\documentclass[aps,prx,twocolumn,superscriptaddress,showpacs,longbibliography]{revtex4-1}
\usepackage[colorlinks=true,citecolor=blue,linkcolor=blue,breaklinks=true]{hyperref}
\usepackage{amssymb}
\usepackage{graphicx}
\usepackage{amsmath}
\usepackage{mathtools}
\usepackage[export]{adjustbox}
\usepackage{epsfig}
\usepackage{times}
\usepackage{xcolor}
\usepackage{subfigure}
\usepackage{setspace}
\usepackage{bm}
\usepackage{calc}
\usepackage{natbib}
\usepackage[normalem]{ulem}
\usepackage{cancel}
\usepackage{xcolor}
\definecolor{darkblue}{RGB}{0,0,150}
\definecolor{nightblue}{RGB}{0,0,100}
\usepackage[normalem]{ulem} 
\usepackage{braket}

\usepackage{cleveref}
\crefname{appendix}{Appendix}{Appendices}

\newcommand{\refsub}[2]{\hyperref[#1]{\ref*{#1}#2}}

\usepackage[english]{babel}
\usepackage[babel,kerning=true,spacing=true]{microtype}

\definecolor{DarkRed}{RGB}{100,0,0}
\definecolor{LightGreen}{RGB}{000,50,0}

\def\({\left(}
\def\){\right)}

\newcommand{\beq}{\begin{equation}}
\newcommand{\eeq}{\end{equation}}
\newcommand{\bal}{\begin{aligned}}
\newcommand{\eal}{\end{aligned}}

\begin{document}
\title{
Theory of post-selected entanglement transitions in monitored bosons
}

\author{Ilia Komissarov}
\email{ilia.komissarov@lsu.edu}
\affiliation{Department of Physics and Astronomy, Louisiana State University, Baton Rouge, Louisiana 70803, USA}
\author{Emanuele G. Dalla Torre}
\affiliation{Department of Physics, Bar-Ilan University, Ramat Gan 52900, Israel}
\author{Ahana Chakraborty}
\email{ahanachakraborty@lsu.edu}
\affiliation{Department of Physics and Astronomy, Louisiana State University, Baton Rouge, Louisiana 70803, USA}

\date{\today}

\begin{abstract}
Entanglement phase transitions driven by quantum measurements have emerged as a central paradigm in open quantum many-body physics. Such phase transitions are well established for
systems with finite local Hilbert-space dimensions, such as qubits and fermions, while their realization in bosonic systems with unbounded local occupation numbers remains poorly understood. Even in the absence of interactions, number states of bosons are intrinsically non-Gaussian, preventing the use of standard correlation-matrix approaches. To address this problem, we develop a replica-free Keldysh field-theoretic framework that expresses the Rényi entropy of bosonic systems initialized in on-site Fock states in terms of permanents of matrices constructed from single-particle Green's functions. Applying this framework to a continuously monitored one-dimensional cross-stitch lattice conditioned on the no-click trajectory, we uncover a transition from volume-law to logarithmic entanglement scaling. We show that the transition is controlled by a restructuring of the non-Hermitian spectrum that changes the number of long-lived modes from extensive to finite. In the strongly monitored regime, bosons dynamically condense into a microscopic number of slowest-decaying modes, producing logarithmic entanglement scaling, whereas an extensive manifold of long-lived modes at weak monitoring gives rise to volume-law entanglement. Our results establish a distinct mechanism for measurement-induced entanglement transitions in free bosonic systems and provide a computationally efficient diagnostic of the measurement-induced bosonic condensation.
\end{abstract}
\maketitle

\section{Introduction} 
The interplay between unitary evolution and measurement in quantum many-body systems has attracted considerable interest, particularly in the context of entanglement behavior \cite{Nahum2021, li2025measurement}. Although a unified theoretical framework describing the entanglement in such systems remains elusive, substantial progress has been made in several paradigmatic settings. One prominent example is monitored free fermions in one dimension, whose entanglement dynamics have been extensively characterized using both analytical \cite{Zhang2022, Kells2023, LeGal2023, Soares2025, Guo2025, muller2022measurement, Cao2019} and numerical approaches \cite{liao2026, bf}. The associated replica Keldysh field theory maps onto a nonlinear sigma model in the AIII Altland--Zirnbauer symmetry class \cite{az}, predicting an area-law entanglement scaling for arbitrary monitoring strength \cite{poboy, fava}, i.e., $S \sim L^{D-1}$, where $S$ is the entanglement entropy, $L$ is the linear system size, and $D=1$
is the system's dimensionality. Upon introducing interactions or non-Gaussian measurements \cite{Lumia}, the behavior becomes significantly richer, allowing for volume-law entanglement ($S \sim L^D$) and measurement-induced phase transitions (MIPT) into area-law phases \cite{Poboiko_2025, Guo2025}. A similar phenomenology was observed extensively in hybrid unitary-measurement circuits of qubits \cite{sm,Skinner2019, Li2018, Choi2020, Gullans2020}. Such qubit-based monitored systems have also been realized experimentally on several conventional quantum computing platforms \cite{Koh2023MIPT,Hoke2023,Noel2022}.

Both aforementioned examples, fermions and qubits, share the same local Hilbert space dimension, $d=2$.
Extending these studies to larger, finite or infinite $d$ has become increasingly important, thanks to both the prospect of qualitatively new theoretical phenomena and recent experimental advances. On the experimental side, bosonic platforms \cite{knill2001scheme, schuster2007circuit, blais2021circuit}, such as multi-mode superconducting cavities coupled to nonlinear elements \cite{schuster2007circuit}, combine long coherence times \cite{reagor_quantum_2016,milul_superconducting_2023}, flexible connectivity \cite{heeres2017implementing,rosenblum2018cnot,chakram_seamless_2021,chakram_multimode_2022,gao2019entanglement,chapman_high--off-ratio_2023}, and the capability for efficient local mid-circuit measurements with tunability in local measurement strength, from parity to number-resolved measurements; all of which are essential for the precise characterization and control of non-unitary quantum dynamics. These advances have motivated growing interest in bosonic quantum-computing platforms and their monitored dynamics.

From a theoretical perspective, it is now well established that measurement-induced transitions in finite-$d$ systems can differ qualitatively from those in the $d\to\infty$ limit. For example, generic finite-$d$ monitored circuits exhibit nontrivial interacting critical points \cite{sm,Aidan_PRL}, whereas certain large-$d$ limits are governed by classical percolation universality \cite{Skinner2019}. Motivated by this distinction, recent work has begun exploring monitored bosonic systems, which are naturally characterized by an unbounded local Hilbert space dimension $d\to\infty$. For a broad class of Gaussian bosonic models, measurements generically dominate the long-time dynamics, leading to area-law entanglement for arbitrary monitoring strength \cite{Minoguchi2022, Yokomizo2025, Barberena2025, Tianciboson}. By analogy with the fermionic case, however, one may expect that sufficiently strong non-Gaussianity could counteract the disentangling effect of measurements and stabilize a genuine measurement-induced transition. This expectation was recently confirmed in interacting bosonic circuits with parity measurements \cite{patel2026}.

Interactions are not the only mechanism capable of generating complex non-Gaussian behavior in bosonic systems, essential to preserve the volume-law-scaling entanglement under monitoring. Unlike fermionic Slater determinants, bosonic Fock states are generically non-Gaussian: i.e., they violate Wick's theorem. For example, for a single free boson state $\hat b^\dagger \ket{0} = |1\rangle$, one finds $\langle1| \hat b^\dagger \hat b^\dagger \hat b \hat b|1\rangle =0$, whereas Wick factorization would incorrectly predict $2\langle1|\hat b^\dagger \hat b |1\rangle^2 =2$ for the same correlator.

Owing to this feature, even non-interacting bosonic dynamics initialized in local Fock states exhibit intrinsically many-body-like entanglement structures. The entanglement entropy generated from initial on-site Fock states can be expressed as a matrix permanent constructed from single-particle wavefunctions \cite{kaga_2023, Kaneko_2025}. While the single-particle dynamics of free systems of bosons is easily solvable, the resulting entanglement structure is highly nontrivial: permanents can be evaluated exactly only for relatively small system sizes, resembling correlators in a genuine many-body system. Using known bounds on permanents \cite{berkowitz2018}, Ref. \cite{kaga_2023} argued that in Hermitian systems, free bosons initialized in local Fock states generically develop volume-law entanglement.

Given this qualitatively distinct entanglement structure of bosons compared with both free fermions and quantum circuits of qubits, a natural question is whether free bosonic systems initialized in local Fock states can likewise exhibit a measurement-induced entanglement transition. In this work, we answer this question affirmatively. Our work advances the understanding of entanglement dynamics in monitored bosonic systems on two fronts: (a) we develop a replica-free non-equilibrium field-theoretic formalism that can treat arbitrary non-Hermitian non-interacting bosonic systems, initialized in non-Gaussian density matrices, especially efficient for computing entanglement in bosonic systems initialized in on-site Fock states with arbitrary occupation. (b) We identify a monitored bosonic model which exhibits a transition from volume law to logarithmic entanglement scaling ($S \sim \log L$) above a critical measurement rate, thus providing an experimentally realizable protocol to observe MIPT in \textit{free} bosonic systems.

On the technical side, we obtain closed-form expressions for the second R\'enyi entanglement entropy by performing a double-branch Keldysh path integral. We use the formalism of the Wigner characteristic function \cite{CG,WCF2,WCF3} to bypass the complexity of replicated contour geometry typically required for computing the R\'enyi entropy \cite{Calabrese_2013}. We then incorporate the non-Gaussianity of the initial state through a derivative construction introduced in Ref.~\cite{Chakraborty_2019}, allowing the entanglement entropy to be represented as differential operators acting on an effective bosonic partition function. This approach efficiently resums the combinatorial complexity of bosonic exchange processes, leading to a general expression for the R\'enyi entropy in terms of permanents of matrices constructed from single-particle Green's functions. Thus, our formalism generalizes previous studies on entanglement in Hermitian systems with singly occupied sites \cite{kaga_2023, Kaneko_2025} to non-Hermitian systems with arbitrary (possibly degenerate) initial occupation. This approach enables us to study bosonic MIPT via a replica-free, analytically tractable field-theoretic approach.

As a direct implementation of our formalism, we focus on monitored systems governed by effective non-stochastic non-Hermitian Hamiltonians \cite{Cao2019, Alberton2021, Biella2021, Gopalakrishnan2021, Zhang2022, Kells2023, Fleckenstein2022, LeGal2023, Zerba2023, Turkeshi2023, Turkeshi2021, Kawabata2023}. Such non-Hermitian systems arise naturally in quantum-jump dynamics upon postselecting the no-jump (post-selected) trajectory, corresponding to the absence of particle-loss events, modeled via imaginary on-site potentials. The non-Hermitian monitored dynamics has been previously studied in fermionic and qubit 
systems, establishing a relation between entanglement transitions and changes in the structure of the non-Hermitian spectrum, including subradiance transitions and the onset of the non-Hermitian skin effect \cite{LeGal2023, Kawabata2023, Fleckenstein2022, Song2019}. Notably, the dynamics of free bosons in such settings is expected to differ essentially from that of free fermions, even when the underlying non-Hermitian Hamiltonian is identical, due to the different statistics of the many-particle wavefunction. In this work, we show that the introduction of imaginary on-site potentials in free bosonic systems can induce Bose-Einstein condensation in a microscopic number of modes, a phenomenon that is impossible in previously studied fermionic systems. 

\begin{figure}[t!tbp]
    \centering
    \includegraphics[width=0.43\textwidth]{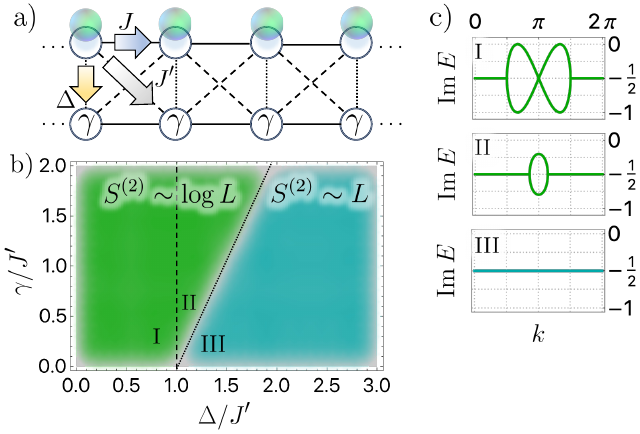}
    \caption{(a) We consider non-interacting bosons hopping on a 1D cross-stitch lattice described by the Hamiltonian \eqref{dimmodel} and initialized in a Fock state with the top (A) sublattice singly occupied by bosons represented by the spheres. The bottom (B) sublattice hosts constant imaginary on-site potentials $\gamma$, modeling the post-selected no-click trajectory of the quantum jump protocol \eqref{psitnc}. (b) The phase diagram exhibits two phases distinguished by the scaling of the steady-state entanglement. The phase with logarithmic scaling of the late-time bipartite second R\'enyi entropy $S^{(2)}_{L/2}$ \eqref{sdef} is shown in green (regions I and II), while the volume-law phase with $S^{(2)}_{L/2} = O(L)$ is shown in blue (region III). The dotted line marks the transition associated with the emergence of an exceptional point of the Hamiltonian $H_{\rm eff}(k)$ at $k=\pi$. The dashed line indicates a subradiant transition that does not affect the scaling of $S^{(2)}_{L/2}$. (c) Representative imaginary spectra ${\rm Im}\,E(k)$ of $H_{\rm eff}(k)$ \eqref{dimmodel} in the three possible parameter regimes. The dispersive nature of ${\rm Im}\,E(k)$ in the regions I, II is responsible for the log-volume scaling of the late-time entanglement, due to the dynamical collapse of all bosons into the small number of slowest-decaying modes. The spectra are computed for $\gamma/J' = 1$ and $\Delta/J' = 0.5$ (I), $1.4$ (II), and $2$ (III).}
    \label{fig1}
\end{figure}

The specific monitored bosonic hopping model we consider is represented by a one-dimensional cross-stitch lattice [see Fig.~\ref{fig1}(a)], where only the bottom sublattice is monitored with the post-selection of the no-click trajectory. We show that this system undergoes a measurement-induced transition from the volume to the log-volume scaling entanglement entropy as the monitoring rate $\gamma$ is varied [see Fig.~\ref{fig1}(b)]. We relate this transition to a qualitative restructuring of the imaginary part of the non-Hermitian spectrum, which determines the number of modes retaining finite occupation at long times. In phases I and II [
Fig.~\ref{fig1}(b,c)], only a single mode remains occupied at late times in the system with open boundaries. Bosons therefore condense into this mode, producing an entanglement entropy that scales as a logarithm of the system size $L$. By contrast, in phase III, an extensive manifold of long-lived modes survives, leading to volume-law entanglement.

This example suggests that the dimensionality of the subspace of stationary modes determines the scaling of the Green's functions and, consequently, the asymptotic scaling of the bosonic permanents that encode entanglement. To generalize this result, we introduce a quantity that diagnoses measurement-induced phase transitions in polynomial time with respect to $L$, thereby bypassing the evaluation of permanents, whose computation scales exponentially with $L$. The roadmap of our work, together with the summary of the main results, is provided below.

\section{Summary of the main results}

\begin{enumerate}
\renewcommand{\labelenumi}{(\arabic{enumi})}

\item We develop a replica-free Keldysh-field-theoretic framework for computing entanglement entropy in a general system of non-interacting bosons evolving under arbitrary non-Hermitian dynamics, including systems describing post-selected no-click measurement trajectories \footnote{In this work, we focus on systems without anomalous (pairing) terms, although such Hamiltonians can be treated analogously using the outlined methods.}. We consider both thermal and on-site Fock initial states. For the latter, the state of the system is non-Gaussian. A key outcome of our formalism is a closed-form expression for the entropy in such systems, allowing us to predict its finite-size scaling analytically. We show that the entanglement entropy in this case develops a many-body-like structure which can be expressed in terms of permanents of matrices constructed from the single-particle Green’s functions. 

\item For such systems initialized in on-site Fock states, we show that different entanglement scaling regimes emerge depending on the number of long-lived modes in the non-Hermitian spectrum. When $O(L)$ modes contribute, the entropy scales with volume. When only a small subset of modes survives at late times, the scaling of the entropy is logarithmic $S \sim \log L$. In this regime, the Green’s function contractions entering the permanents that define entanglement entropy become effectively rank-1, with $S$ admitting a simplified analytic form controlled by the spatial support of the surviving modes. 

\item As a concrete realization of the measurement-induced transition between these two regimes, we introduce a one-dimensional dimer chain model with continuous boson number monitoring of a subset of sites and analyze the no-click quantum trajectory. The effective non-Hermitian Hamiltonian exhibits momentum-resolved exceptional point transitions, which reorganize the imaginary part of the spectrum. As the monitoring rate and inter-sublattice coupling are varied, the system transitions from a regime in which all modes decay uniformly to one in which only a small subset of modes dominates the long-time dynamics, giving rise to a transition between volume and log-volume scaling behavior of the bipartite entanglement. A key finding of our work is the introduction of a polynomial-time diagnostic for this transition based on the rank structure of the overlap matrices of the single-particle Green's function, which allows us to avoid the exponential complexity of evaluating permanents.

\end{enumerate}
The remainder of this paper is organized as follows. In Sec.~\ref{subsec:no-click}, we first introduce no-click trajectories as a concrete realization of non-Hermitian dynamics and discuss the general structure of entanglement in monitored bosonic systems initialized in Fock states in Sec.~\ref{subsec:struct}. In \crefrange{sec:form}{subsec:FockIn}, we develop a replica-free Keldysh field-theoretic framework for computing the Rényi entropy of non-Hermitian bosonic systems and derive general expressions for both thermal and on-site Fock initial states. We then analyze the entanglement structure of monitored bosons and relate the scaling of the entropy to the properties of single-particle Green's functions and overlap matrices in Sec.~\ref{sscaling}. In~\crefrange{subsec:onedimer}{subsec:dimerlattice}, we introduce a monitored cross-stitch lattice and demonstrate a measurement-induced transition from volume-law to logarithmic entanglement scaling, which we connect to a restructuring of the non-Hermitian spectrum and dynamical condensation into long-lived modes. We further discuss the dependence of the transition on the initial state in subsection \ref{subSec:thermNoclick}, arguing that the measurement-induced mode condensation transition for the Fock state is replaced by a dynamical purification transition if the system is instead initialized in a thermal density matrix.

\section{Entanglement entropy of bosons in a general non-Hermitian system}

\label{sec3}

In this section, we develop a general framework for calculating the second R\'enyi entropy of non-interacting bosons evolving under non-Hermitian dynamics. After introducing no-click trajectories as a paradigmatic realization, we formulate the entanglement entropy in terms of the Wigner characteristic function and compute it for both thermal and on-site Fock initial states. For Fock initial states, the effective state of the system is non-Gaussian, and this construction yields an exact representation of the entanglement in terms of matrix permanents, revealing the relation between long-time entanglement scaling and long-time behavior of single-particle Green's functions.

\subsection{No-click trajectory as a concrete example of non-Hermitian Hamiltonian evolution}\label{subsec:no-click}

In this work, we focus on the dynamics of bosonic systems subject to unitary evolution interspersed with measurements. The unitary dynamics is generated by a Hamiltonian $\hat H$, while the measurements are described by Kraus operators $\hat K_\tau$ acting at discrete times $\tau \delta_t$, where $\tau=1,\ldots,T$, and $\delta_t$ is an infinitesimal time step. The resulting quantum trajectories are
\begin{equation}
|\psi(t)\rangle=
\hat K_T e^{-i\hat H\delta_t}\cdots
\hat K_2 e^{-i\hat H\delta_t}
\hat K_1 e^{-i\hat H\delta_t}
|\psi(0)\rangle \, ,
\label{psitk}
\end{equation}
with $t=T \delta_t$. Throughout this work, we focus on the correlation functions and entanglement within individual quantum trajectories, rather than their averages over measurement outcomes. We focus on measurements for which the Kraus operators can be written as
\begin{equation}
\hat K=e^{\hat{\mathcal O}\delta_t},
\label{expform}
\end{equation}
where $\hat{\mathcal O}$ is an operator at most quadratic in bosonic creation and annihilation operators $\hat b^\dagger$, $\hat b$. In this case, a Gaussian initial state remains Gaussian throughout the non-unitary evolution, allowing for an efficient analytical description. 

A paradigmatic example is provided by the no-click trajectory within the quantum jump protocol
\cite{Turkeshi2021} where loss events on the site $x$ are described by the jump operators $\hat K_{x}^{(1)} = \sqrt{2\gamma \delta_t}\hat b_{x}$ with $\gamma$ denoting the measurement rate. The Kraus operator describing the absence of loss events at a site $x$ is
\begin{equation}
\hat K_{x}^{(0)}
=
1-\gamma\delta_t\,\hat b_x^\dagger \hat b_x
\simeq
e^{-\gamma\delta_t\,\hat b_x^\dagger \hat b_x}\, ,
\label{nc}
\end{equation}
assuming the form \eqref{expform}. If no loss events occur  at any of the sites $x$, throughout the entire evolution \eqref{psitk}, taking the continuum limit $\delta_t\to0$ while keeping $t=T \delta_t$ fixed, yields
\begin{equation}
|\psi(t)\rangle
=
e^{-i\left(\hat H-i\gamma\sum_x \hat b_x^\dagger \hat b_x\right)t}
|\psi(0)\rangle,
\label{psitnc}
\end{equation}
implying that the no-click trajectory is governed by the effective non-Hermitian Hamiltonian
\begin{equation}
\hat H_{\rm eff}
=
\hat H-i\gamma\sum_x \hat b_x^\dagger \hat b_x.
\label{Heff}
\end{equation}
Since $\hat H_{\rm eff} \neq \hat H_{\rm eff}^\dagger$, the norm of the state is not conserved. Consequently, expectation values of operators $\hat {\cal O}$ in the non-Hermitian system state density matrix $\hat \rho(t)$ need to be appropriately normalized at any $t>0$,
\begin{equation}
\braket{\hat {\cal O}}
=
\frac{{\rm Tr}[\hat \rho(t) \, \hat {\cal O}]}
{{\rm Tr}\, \hat \rho(t)}\, ,
\label{norm}
\end{equation}
where the denominator is the probability of the corresponding measurement record. Another important example of a non-Hermitian effective Hamiltonian arising in monitored quantum systems is the Hatano--Nelson Hamiltonian obtained when a system is continuously monitored via local particle-current measurements, conditioned on a finite directed current \cite{Kawabata2023}, which can be analyzed using the same approach via our formalism introduced below.

\subsection{Entanglement structure of free bosons initialized in on-site Fock states}\label{subsec:struct}

In the remainder of this section, we describe qualitatively the entanglement dynamics of bosons initialized on the sites and evolved under an arbitrary non-interacting non-Hermitian Hamiltonian $\hat H_{\rm eff}$. This includes, but is not restricted to, the no-click trajectory evolution with~\eqref{Heff}. We quantify the entanglement through the second R\'enyi entanglement entropy of a subsystem $\rm s$, normalized according to \eqref{norm}, defined as
\begin{equation}
S^{(2)}_{\rm s}(t)=-\log\left[\frac{{\rm Tr_{\rm s}}\left[\left({\rm Tr}_{\overline{\rm s}}\,\hat\rho(t)\right)^2\right]}{\left[{\rm Tr}\,\hat\rho(t)\right]^2}\right]\, .
\label{sdef}
\end{equation}
We will especially be interested in the bipartite entropy $S^{(2)}_{L/2}$, where ${\rm s}=L/2$ is half of the system in real space.

\begin{figure*}[t!tbp]
    \centering
    \includegraphics[width=0.9\textwidth]{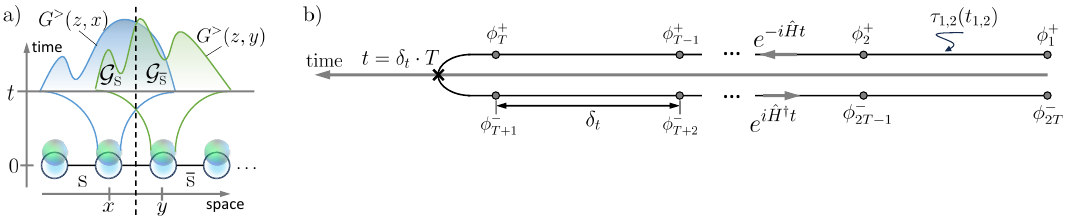}
    \caption{a) Initially localized bosons, represented by green spheres, are evolved using the Green's function $\hat G^>$, forming extended probability density clouds, shown in blue and green spatial profiles. The overlaps ${\cal G}_{\rm s}(x,y)$, ${\cal G}_{\rm \overline s}(x,y)$ (see \eqref{calgs}) of the wave functions to the left and to the right of the bipartition, indicated with the dashed line, give rise to the finite entanglement entropy in the system \eqref{SNNmain}. The entropy is maximized when the overlaps are equal on both sides of the bipartition, thereby maximizing the uncertainty in particle positions. b) Keldysh time contour corresponding to the discretization of the time evolution of the density matrix $\hat \rho(t) = e^{i \hat H_{\rm eff}^\dagger t} \hat \rho(0) e^{-i \hat H_{\rm eff} t}$ at regular time intervals with spacing $\delta_t$. The bosonic fields living on the forward branch of the contour are labeled as $\phi^+_\tau$, and on the backward $\phi^-_\tau$ with $\tau$ varying from 1 to $2T$. Due to the non-Hermiticity of the Hamiltonian $\hat H_{\rm eff}$, the contour cannot be extended to $+\infty$, and instead terminates at a definite time $t = T \cdot \delta_t$ indicated by a cross.}
    \label{fig2}
\end{figure*}

Before presenting the formal derivation, we briefly discuss the physical origin and structure of entanglement in non-Gaussian bosonic systems evolving under free non-Hermitian dynamics. For concreteness, consider an initial Fock state containing one boson on each site of a one-dimensional hopping model, as illustrated in \autoref{fig2}a,
\begin{equation}
|\psi(0)\rangle=\prod_{x}\hat b_x^\dagger|0\rangle.
\label{Fock}
\end{equation}
Since this state is a direct product of localized single-particle states, its bipartite entanglement entropy \eqref{sdef} vanishes initially.

The time evolution of a single boson state initially localized at site $x$ from $t_0 = 0$ to a finite time $t>0$ is determined by the single-particle vacuum Green's function 
\begin{equation}
\hat b_x^\dagger |0\rangle\rightarrow\sum_z G^{>}(z,x)\hat b_z^\dagger |0\rangle\, ,
\label{gf}
\end{equation}
where
\begin{equation}
G^{>}(z,x)=\langle0|\hat b_z e^{-i\hat H_{\rm eff}t}\hat b_x^\dagger|0\rangle=\sum_E\psi_E^{\rm R}(z)\psi_E^{{\rm L}*}(x)e^{-iEt},
\label{gg}
\end{equation}
and $\psi_E^{\rm R}(x) = \braket{x|E^{\rm R}}$, $\psi_E^{\rm L}(x) = \braket{x|E^{\rm L}}$ denote the right and left wavefunctions of the non-Hermitian Hamiltonian, with
\begin{equation}
\hat H_{\rm eff}|E^{\rm R}\rangle=E|E^{\rm R}\rangle,\qquad \hat H_{\rm eff}^\dagger|E^{\rm L}\rangle=E^*|E^{\rm L}\rangle \, .
\label{ERL}
\end{equation}

As follows from the definition \eqref{gg}, $G^{>}(z,x)$ is simply the wavefunction at time $t$ and position $z$ of the particle initialized at the site $x$. As time progresses, the initially localized bosons (represented by spheres in \autoref{fig2}a) spread into states characterized by extended single-particle wavefunctions, illustrated by the blue and green profiles in \autoref{fig2}a. Whenever these wavefunctions have support on both the region $\rm s$ and its complement $\overline{\rm s}$, tracing out one subsystem produces a mixed reduced density matrix, resulting in a finite entanglement entropy \eqref{sdef}. This contribution to entanglement is encoded in the diagonal ($x=y$) components of the subsystem-resolved overlap matrix
\begin{equation}
{\cal G}_{\rm s(\overline s)}(x,y)=\sum_{z\in{\rm s}(\overline {\rm s})}G^{>\dagger}(x,z)G^{>}(z,y)\, .
\label{calgs}
\end{equation}
These diagonal contributions signify the probability of the boson starting at the site at $x=y$ to propagate into the region $\rm s(\overline s)$, with the entanglement maximized when ${\cal G}_{\rm s}(x,x) \simeq {\cal G}_{\rm \overline{s}}(x,x)$, due to the maximal positional uncertainty with respect to the bipartition.

For many-boson states, additional contributions arise from bosonic exchange processes determined by the overlaps of the wavefunctions of different bosons ${\cal G}_{\rm s}(x,y)$, $x \neq y$ (see \autoref{fig2}a). Thus, entanglement acquires a complicated many-boson combinatorial structure involving coherent symmetrization over all bosonic exchange processes determined by \eqref{calgs}, ultimately giving rise to the bosonic permanents of $\hat {\cal G}_{\rm s}$ and related quantities, as shown below.

While the discussion above applies equally well to Hermitian systems, the Green's functions \eqref{gg} may exhibit qualitatively new behavior in the non-Hermitian setting. As one example, in systems with the non-Hermitian skin effect, the late-time Green's functions acquire a strongly non-reciprocal character, $|G^>(x,y)| \gg |G^>(y,x)|$, due to the dynamical build-up of modes on one boundary of the system, which is reflected in the area-law scaling behavior of the steady-state entanglement of fermions \cite{Kawabata2023}. Another essential difference arises from the presence of the imaginary part of the spectrum, ${\rm Im}\,E \neq 0$. In monitored systems, ${\rm Im}\,E$ can become dispersive, leading to a time-dependent spectral filtering of modes (see \eqref{gg}). As we demonstrate below, this feature can, in some cases, give rise to a volume-to-log-volume entanglement scaling transition in bosonic systems, due to the collapse of all particles into a small subset of long-lived modes.

To analytically compute the entanglement entropy \eqref{sdef}, we utilize the formalism of the Wigner characteristic function, which we review below. We then determine the entanglement entropy in systems with thermal initial conditions. These systems retain Gaussianity under evolution with a non-interacting Hamiltonian, allowing us to benchmark our results against the existing work on monitored Gaussian systems. Then, we proceed to treat on-site Fock initial conditions that violate Gaussianity, which requires modifications to existing approaches.

\subsection{Entanglement entropy from the Wigner characteristic function in general non-Hermitian bosonic systems}
\label{sec:form}

Commonly used field-theoretic approaches for computing Rényi entropies involve evaluating correlation functions on replicated space-time manifolds containing branch cuts along the subsystem defining the reduced density matrix \cite{Calabrese_2013}. While powerful, these methods are often limited by the complicated boundary conditions imposed in the replicated geometry. Here, we circumvent this difficulty using a Wigner characteristic function (WCF) approach, which enables the computation of the second Rényi entropy in a generic bosonic system without introducing replicated geometries \cite{Chakraborty_2021}. We briefly review the WCF formalism below and, in the next subsection, show how both the WCF and the Rényi entropy can be obtained from a single replica-free path-integral formulation.

The WCF is defined via the expectation value of the displacement operator $\hat D = e^{\sum_x \left(\xi_x \hat b_x^\dagger - \xi_x^* \hat b_x\right)}$ \cite{CG, ACN}, as
\begin{equation}
\chi^{\rm W}(t,\{\xi_x\}) = {\rm Tr}\!\left[\hat\rho(t)e^{\sum_x \left(\xi_x \hat b_x^\dagger - \xi_x^* \hat b_x\right)}\right]\, ,
\label{WCFd}
\end{equation}
where $\{\xi_1, \xi_1^*, \ldots, \xi_L, \xi_L^* \}$ can be loosely thought of as the phase variables. The Wigner characteristic function can be used to obtain the second R\'enyi entropy, eq.~\eqref{sdef} (see Appendix \ref{appb} for a full derivation):
\begin{equation}
e^{-S_{\rm s}^{(2)}(t)}
=
\frac{
\displaystyle
\int \left\{
\prod_{y\in{\rm s}}
\frac{d\xi_y\, d\xi_y^*}{\pi}
\right\}
\chi^{\rm W}(t,\{-\xi_y\})\chi^{\rm W}(t,\{\xi_y\})
}{
\left[{\rm Tr} \,\hat\rho(t) \right]^2
} \, ,
\label{s2main}
\end{equation}
where the denominator results from the normalization convention \eqref{norm}. As illustrated in Ref. \cite{Chakraborty_2021}, and follows from \eqref{WCFd}, each WCF can be expressed as a path integral of the partition function ${\rm Tr}\,\hat \rho(t)$ with additional source terms turned on at time $t$ within the subsystem $x\in {\rm s}$. Thus, the integration over $\{\xi_x\}$ in \eqref{s2main} is equivalent to the conventional tracing over the subsystem in \eqref{sdef}. 

In the next subsection, we suitably modify this approach to incorporate the effect of measurement back-reaction during the dynamics. Once $\chi^{\rm W}(t,\{\xi_x\})$ is determined, the entanglement entropy of the non-Hermitian system follows directly from \eqref{s2main}. We show how this formalism applies to two different choices of initial conditions: a system of free bosons initially prepared in a thermal state and in the local Fock state.

\subsection{Entanglement entropy initialized in a thermal density matrix}\label{subsec:thermIn}

As a warm-up and to benchmark against the existing approaches to entanglement in monitored systems, we first consider bosonic systems initialized in thermal states. For non-Hermitian systems of free bosons with a Gaussian initial condition, such as the thermal density matrix, the WCF \eqref{WCFd} can be evaluated exactly using the Keldysh contour, sketched in \autoref{fig2}b. To that goal, we introduce on-site many-boson coherent states $\ket{\phi_\tau(1),\ldots,\phi_\tau(L)}$ satisfying
\begin{equation}
\hat b_x\ket{\phi_\tau(1),\ldots,\phi_\tau(L)}=\phi_\tau(x)\ket{\phi_\tau(1),\ldots,\phi_\tau(L)} \, ,
\end{equation}
where we assume a discrete system in the real space with $L$ sites. As noted in earlier works on monitored systems \cite{poboy, yangkeldysh} and in contrast to the Hermitian case \cite{Chakraborty_2021}, the Keldysh contour cannot be extended to an arbitrary final time, usually set equal to $t_{\infty} = +\infty$, and must terminate at the finite physical time $t$, due to the decay of the normalization of the state with time.

With this subtlety taken into account, the time-discretized coherent-state path-integral representation of the Wigner characteristic function \eqref{WCFd} in an arbitrary bosonic system takes the form
\begin{equation}
\begin{aligned}
\chi^{\rm W}(t,\{\xi_x\})
& =\int D\phi^+ D\phi^-\,
e^{i[S^+(t)-S^-(t)]} \\
& \times 
e^{i\sum_{x=1}^L\left[\xi_x\phi_T^{+*}(x)-\xi_x\phi_{T+1}^{-*}(x)+{\rm h.c.}\right]} \, ,
\end{aligned}
\label{chiRRspm}
\end{equation}
where we considered Gaussian actions on the forward ($\phi^+$) and backward ($\phi^-$) branches of the Keldysh contour. In a non-interacting system of bosons, the integration measure is defined as
\begin{equation}
\begin{gathered}
S^{+}(t)
= \sum_{\tau = 1}^{T} \sum_{x,y=1}^L\,
\phi_\tau^{+ *}(x)
\bigl[i\partial_{\tau}\delta_{xy}-H^{+}(x,y)\bigr]
\phi_
\tau^{+}(y) \, , \\
 S^{-}(t)
= \sum_{\tau = T+1}^{2T} \sum_{x,y=1}^L\,
\phi_\tau^{- *}(x)
\bigl[i\partial_{\tau}\delta_{xy}-H^{-}(x,y)\bigr]
\phi_
\tau^{-}(y) \, , \\
D\phi^{+}= \prod_{\tau=1}^{T}\prod_{y=1}^L\frac{d\phi_\tau^{+ *}(y)\,d\phi_\tau^{+}(y)}{\pi} \, , \\ D\phi^{-}= \prod_{\tau=T+1}^{2T}\prod_{y=1}^L\frac{d\phi_\tau^{- *}(y)\,d\phi_\tau^{-}(y)}{\pi} \, ,
\end{gathered}
\label{spm}
\end{equation}
with $H^{+}(x,y)=\braket{x|\hat H|y}$,
$H^{-}(x,y)=\braket{x|\hat H^\dagger|y}$, $\ket{x}=\hat b_x^\dagger\ket{0}$.
The integral \eqref{chiRRspm} is a familiar discretization of a partition function ${\rm Tr} \, \hat \rho$ \cite{kam} with the addition of instantaneous $\{\xi_x\}$ sources entering \eqref{WCFd} turned on only at the time $t$ at which the WCF $\chi^{\rm W}(t,\{\xi_x\})$ is computed, which coincides with the time of the terminus of the Keldysh contour (\autoref{fig2}b).

After performing the Keldysh rotation
$\phi_\tau^{\rm cl}(x)=(\phi_\tau^+(x) + \phi_\tau^-(x))/\sqrt{2}$,
$\phi^{\rm q}_\tau(x)=(\phi^+_\tau(x) - \phi^-_\tau(x))/\sqrt{2}$,
Eq.~\eqref{chiRRspm} can be written as a Keldysh partition function with sources $\xi_x$
coupled to $\phi_{\rm cl}$ at time $t$ \cite{Chakraborty_2021}. If the considered actions $S^\pm$ are Gaussian, as in \eqref{spm}, the integrals in \eqref{chiRRspm} can be evaluated exactly, in the continuum limit $\delta_t \to 0$,  with a finite $t = T \delta_t $, resulting in
\begin{equation}
\chi^{\rm W}(t,\{\xi_x\})
=
{\rm Tr}\,\hat\rho(t)\,
e^{
-\frac{i}{2}\sum_{x,y}\xi_x^* G_{\rm K}(x,y,t,t,t)\,\xi_y
}\,,
\label{WCFgauss}
\end{equation}
where $G_{\rm K}$ is the Keldysh Green's function
\begin{equation}
iG_{\rm K}(x,y,t,t,t) = {\rm Tr} \left[ (\hat b_x^\dagger \hat b_y + \hat b_y \hat b_x^\dagger ) \hat \rho(t) \right]\, .
\label{GK}
\end{equation} 
The first and second time arguments of this quantity correspond to the placement of $\hat b$-operators in the Keldysh contour, while the last argument refers to the contour termination time (see \autoref{fig2}b). Substituting Eq.~\eqref{WCFgauss} into Eq.~\eqref{s2main} and performing the Gaussian integral over the subsystem variables $\{\xi_x\}$ yields the compact determinant formula for the entanglement entropy
\begin{equation}
S_{\rm s}^{(2)}(t)
=
{\rm Tr}\log
\left[
iG_{\rm K}(x,y,t,t,t)
\right]_{x,y \in \rm s} \, .
\label{eq:S2_det_main}
\end{equation}

The expression above is a reflection of the well-known fact that in both bosonic and fermionic Gaussian systems, the entropy is determined exclusively by single-particle correlators, such as the correlation matrix \cite{LeGal2023, Zerba2023, Turkeshi2023, Kawabata2023, Turkeshi2021}
\begin{equation}
C(x,y,t) = {\rm Tr} \left[ \hat b_x^\dagger \hat b_y \hat \rho(t) \right] \, .
\label{corr1}
\end{equation}
Indeed, using \eqref{corr1}, \eqref{GK}, the entanglement entropy \eqref{eq:S2_det_main} can be written as 
\begin{equation}
S_{\rm s}^{(2)}(t) = \sum_{\alpha} \log[1 + 2 \lambda_{\alpha}(t)] \, ,
\label{alphasum}
\end{equation}
where $\lambda_{\alpha}$ are eigenvalues of the correlation matrix \eqref{corr1} restricted to the subsystem $[C(x,y,t)]_{x,y \in \rm s}$. In particular, for a thermal state in the absence of monitoring, the eigenvalues $\lambda_{\alpha}$ are finite, and the sum \eqref{alphasum} receives contributions from $O(L)$ modes, giving rise to the familiar volume-law scaling of the thermal entropy \cite{ACN}. We apply the above approach and use \eqref{eq:S2_det_main} to study entanglement phases of monitored bosons initialized in a thermal density matrix and post-selected to no-click trajectories in subsection \ref{subSec:thermNoclick}.

\subsection{Entanglement entropy for initial Fock states. }\label{subsec:FockIn}

We now consider the case where the system is initialized in an arbitrary on-site Fock state with $n_x$ bosons per site $x$,
\begin{align}
\ket{\psi(0)} = \prod_x \frac{(\hat b^\dagger_x)^{n_x}}{\sqrt{n_x!}} \ket{0} \, .
\label{Fock}
\end{align}
For such a state, the WCF can no longer be written as a Gaussian integral over coherent states, and \eqref{eq:S2_det_main} no longer holds. The non-Gaussianity arises from the non-exponentiable nature of the representation of $\ket{\psi(0)}$ in the coherent-state basis. This difficulty can be resolved by utilizing a derivative trick \cite{Chakraborty_2019} 
\begin{equation}
\begin{aligned}
\left\langle \phi_1(x) \middle|n_x\right\rangle 
\left\langle n_x\middle|\phi_{2T}(x)\right\rangle
&=
\frac{\bigl(\bar{\phi}_1(x)\phi_{2T}(x)\bigr)^{n_x}}{n_x!}
 \\  & = 
\left.
\frac{1}{n_x!}\partial_{u_x}^{n_x}
e^{u_x \bar{\phi}_1(x)\phi_{2T}(x)}
\right|_{\{u_x\}=0} \, .
\end{aligned}
\end{equation}
Applying this identity for each initially occupied site, we write the WCF as a Gaussian integral \eqref{chiRRspm} with both forward and backward branch actions depending on auxiliary variables $\{u_x\}$ (see \cref{appe}). Evaluating the integral over the coherent states, we find
\begin{equation}
\begin{aligned}
\chi^{\rm W}(t,\{\xi_x\})
&=\partial_{\{u_x\}}{\rm det}^{-1}\left(\hat I-\hat{\cal G}_{{\rm s}+\overline{\rm s}}\hat\Delta\right)
\\
&\qquad\times e^{-\frac{i}{2}\sum_{x,y}\xi_x^*G^{\rm K}_{xy}(\{u_x\})\xi_y}\Big|_{\{u_x\}=0},
\\
\hat G^{\rm K}(\{u_x\})
&=\hat I+2 \hat G^>\hat \Delta\left(\hat I-\hat{\cal G}_{{\rm s}+\overline{\rm s}}\hat\Delta\right)^{-1}\hat G^{>\dagger}.
\end{aligned}
\label{chifock}
\end{equation}
where $\Delta(x,y) = u_x \delta_{x,y} $ represent the quadratic sources coupled to bilinears of the initial fields, $\partial_{\{u_x\}} = \prod_{x=1}^L (n_x!)^{-1}\partial_{u_x}^{n_x}$ encodes the initial condition and ${\cal G}$ is defined in Eq.~(\ref{calgs}). Note that we use the notation $\hat {\cal O}$ for operators, while their matrix elements are denoted as ${\cal O}(x,y) = \braket{x|\hat {\cal O}|y}$. It is instructive to compare this expression to the WCF in a Gaussian system \eqref{WCFgauss}. The $\det$ prefactor corresponds to ${\rm Tr}\, \hat \rho$ in \eqref{WCFgauss} computed in the presence of sources $\{u_x\}$, while the matrix $\hat G^{\rm K}(\{u_x\})$ is the effective Keldysh Green's function in the presence of $\{u_x\}$. The derivatives in \eqref{chifock} can be evaluated explicitly, yielding closed-form expressions in terms of Laguerre polynomials \cite{Chakraborty_2021} (see \cref{appe} for more details).

For computing the entanglement entropy, it is convenient to postpone the evaluation of derivatives, taking the Gaussian integral over the $\{\xi_x\}$ variables in \eqref{s2main} first, which yields (see \cref{appe})
\begin{equation}
e^{-S_{\mathrm{s}}^{(2)}}=\frac{\left.\partial_{\{u_x,\widetilde{u}_x\}}\operatorname{det}^{-1}\!\left(\begin{array}{cc}\hat I+\hat{\cal G}_{\overline{\mathrm{s}}}\hat\Delta&\hat{\cal G}_{\mathrm{s}}\hat{\widetilde\Delta}\\\hat{\cal G}_{\mathrm{s}}\hat\Delta&\hat I+\hat{\cal G}_{\overline{\mathrm{s}}}\hat{\widetilde\Delta}\end{array}\right)\right|_{\{u_x,\widetilde u_x\}=0}}{\left(\left.\partial_{\{u_x\}}\det^{-1}\!\left(\hat I+\hat{\cal G}_{\mathrm{s}+\overline{\mathrm{s}}}\hat\Delta\right)\right|_{\{u_x\}=0}\right)^2}.
\label{sfock}
\end{equation}
The denominator in this expression is a normalization factor arising from ${\rm Tr}\, \hat \rho$ factors in \eqref{s2main}, while the numerator contains two copies of derivatives corresponding to two WCFs in \eqref{s2main}, and $\widetilde \Delta(x,y) = \widetilde u_x \delta_{x,y}$. Evaluating the derivatives head-on leads to a combinatorial explosion of terms. To organize it, it is convenient to use MacMahon's master theorem \cite{mmt}, expressing inverse determinants in terms of permanents:
\begin{equation}
\operatorname{det}^{-1}(\hat{I}-\hat{M} \hat{\Delta})=\sum_{n_1, \ldots, n_L} \operatorname{perm} \hat{M}\left[\left\{n_x\right\}\right] \prod_x \frac{u_x^{n_x}}{n_x!},
\label{MMT}
\end{equation}
where $ \hat{M}\left[\left\{n_x\right\}\right]$ denotes the matrix obtained from $\hat M$ by repeating its $x$-th row and $x$-th column $n_x$ times, and the permanent is defined as
\begin{equation}
\operatorname{perm} \hat{M}=\sum_{\sigma \in {\cal S}_N} \prod_{i=1}^N M_{i \sigma(i)}\, ,
\label{perm}
\end{equation}
to be compared with
\begin{equation}
\operatorname{det} \hat{M}=\sum_{\sigma \in {\cal S}_N}(-1)^{\operatorname{sgn} \sigma} \prod_{i=1}^N M_{i \sigma(i)}\, ,
\end{equation}
where ${\cal S}_N$ is the set of all permutations of integers from 1 to $N$. MacMahon's master theorem provides a way to efficiently compute the derivatives in both the numerator and the denominator of \eqref{sfock}, thereby obtaining the entanglement entropy of bosons initialized in on-site Fock states in terms of permanents \eqref{perm}. The expression for the entropy \eqref{sfock} together with the prescription to evaluate the derivatives \eqref{MMT} constitute the main result of this work, applicable to any non-Hermitian system of bosons initialized in an arbitrary on-site Fock state.

In this work, we focus on initial conditions in which a subset A of sites is singly occupied (see, e.g., \autoref{fig1}a, where A corresponds to the top row of sites). In this case, from \eqref{sfock}, \eqref{MMT} we find
\begin{equation}
e^{-S_{\rm s}^{(2)}}=
\frac{
\operatorname{perm}
\begin{bmatrix}
{\cal G}_{\mathrm{s}}(x,y)
&
{\cal G}_{\overline{\rm s}}(x,y)
\\
{\cal G}_{\overline{\rm s}}(x,y)
&
{\cal G}_{\rm s}(x,y)
\end{bmatrix}_{x,y \in {\rm A}}
}{
\operatorname{perm}^2 \left[{\cal G}_{{\rm s} + {\overline {\rm s}}} (x,y)\right]_{x,y \in {\rm A}}
} \, ,
\label{SNNmain}
\end{equation}
where the overlap matrices $\hat {\cal G}_{\rm s(\overline \rm s)}$ are defined in \eqref{calgs}, and
\begin{equation}
    {\cal G}_{{\rm s} + {\overline {\rm s}}}(x,y)=\sum_{z}G^{>\dagger}(x,z)G^{>}(z,y) \,.
\end{equation}
A similar expression was obtained in the Hermitian case \cite{kaga_2023, Kaneko_2025}. In these works, however, the denominator was absent, as for Hermitian systems right and left eigenstates are separately orthonormal and ${\cal G}_{{\rm s} + {\overline {\rm s}}}(x,y) = \delta_{x,y} $.

The numerator in \eqref{SNNmain} contains the bosonic wavefunction overlap data resolved with respect to the bipartition, encoded in matrices $\hat{\cal G}$ defined in \eqref{calgs} and illustrated in \autoref{fig2}a. It implies that the entanglement is controlled by two ingredients: how much each bosonic wavefunction is split across the cut, and how strongly different bosonic wavefunctions overlap with one another. The entanglement entropy is enhanced when the wavefunction weight is distributed comparably between the two halves, $\hat {\cal G}_{\rm s} \simeq \hat {\cal G}_{\overline {\rm s}}$, while it vanishes when the wavefunctions remain confined to one side of the bipartitions. An extreme example is when $\hat {\cal G}_{\rm s} = 0$: in this case $\hat {\cal G}_{\rm s + \overline s} = \hat {\cal G}_{\overline s}$, and the denominator in \eqref{SNNmain} cancels the numerator, resulting in $S^{(2)}_{\rm s} = 0$. Also note the symmetry of the expression \eqref{SNNmain} with respect to the exchange of ``s'' and ``$\rm \overline s$'', following from the invariance of the permanents \eqref{perm} with respect to the permutation of rows and columns.

In Refs. \cite{kaga_2023, Kaneko_2025}, it was determined that in Hermitian systems of hopping bosons, the entropy \eqref{SNNmain} scales extensively with the size of the system. Below, we show that this result does not hold in general for non-Hermitian systems, and that monitoring can induce a phase transition to a regime in which the entropy scales logarithmically with the system size. 

\subsection{Entanglement entropy scaling in non-Hermitian bosonic systems}
\label{sscaling}

The back-reaction of measurement in monitored systems generically leads to the decay of certain components of the wavefunction. One example is the no-click post-selection limit \eqref{psitnc}, which forces the modes supported on the monitored sites to decay exponentially in time. In a translationally-invariant system, this often leads to a finite dispersion in the imaginary spectrum ${\rm Im}\, E(k)$ of the model described by a non-Hermitian effective Hamiltonian $\hat H_{\rm eff}$. Here, we show that this feature can give rise to a log-volume ($S^{(2)}_{L/2} \sim \log L$) scaling of the late-time entanglement entropy of the bosonic system initialized on the sites \eqref{SNNmain}. 
 
For simplicity, we assume that the spectrum ${\rm Im}\, E(k)$ has a global maximum, and denote the right and left wavefunctions of the slowest decaying mode as $\chi^{\rm R}(x)$ and $\chi^{\rm L}(x)$. In this case, the overlaps $\hat {\cal G}_{\rm s(\rm \overline s)}$ \eqref{calgs} at late times reduce to rank-1 matrices $\left.{\cal G}_{\rm s(\rm \overline s)} (x,y)\right|_{t \to \infty} \sim \chi^{\rm L}(x) \chi^{\rm L*}(y)$. This allows us to obtain a simplified expression for the entanglement entropy (see \cref{appe}, \eqref{sbinomial})
\begin{equation}
e^{-S_{\rm s}^{(2)}(\infty)}
\simeq
\sum_{k,m=0}^L
\binom{L}{k}\binom{L}{m}\binom{k+m}{k}
\left(-\frac{c_{\rm s}^{\rm R}}{c^{\rm R}}\right)^{k+m} \, ,
\label{slog}
\end{equation}
where the constant $c_{\mathrm{s}}^{\mathrm{R}}=\sum_{x \in \mathrm{s}}\left|\chi^{\mathrm{R}}(x)\right|^2$ denotes the support of the slowest decaying mode in the subsystem $\rm s$, and $c^{\mathrm{R}}=\sum_x\left|\chi^{\mathrm{R}}(x)\right|^2$ is the normalization coefficient different from unity due to the eigenstate non-orthonormality in non-Hermitian systems. In systems invariant under parity $c_{L/2}^{\rm R} = c_{\overline{L/2}}^{\rm R}$, and the late-time bipartite entropy \eqref{slog} exhibits log-volume scaling $S_{L/2}^{(2)}(t \to \infty) = \log \sqrt{\pi L}$. In \cref{appe}, we show that this scaling is motivated by the dynamical condensation of all bosons into the slowest-decaying mode $\chi$. In particular, \eqref{slog} corresponds to the R\'enyi entropy of a Bose-Einstein condensate, as we show by comparing it with the entropy of bosons initialized on a single site, which is also captured by our general derivative expression \eqref{sfock}. This phenomenon is analogous to the dynamical collapse of bosons into a single state observed in non-Hermitian systems by Refs. \cite{np1, np2, np3}, albeit in a different regime.


\subsection{A computationally efficient diagnostic of the entanglement phase transition}
\label{srank}

The realization that the entropy scaling is tied to the rank of the overlap matrices $\hat {\cal G}$ allows us to introduce a computationally efficient diagnostic of the measurement-induced bosonic condensation. This allows us to bypass the evaluation of the permanents in \eqref{SNNmain}, whose computation scales exponentially with system size. We define the effective rank
\begin{equation}
{\rm rank}\, \hat {\cal G}_{\rm s}(t) = \frac{\sum_i |\sigma_i(t)|}{{\rm max} \, |\sigma_i(t)|} \, ,
\label{rk}
\end{equation}
where $\sigma_i(t)$ are the singular values of the $L \times L$ matrix $\hat {\cal G}{\rm s}$, which can be obtained in polynomial time in the system size. The quantity \eqref{rk} measures the number of modes that effectively contribute at time $t$: we conjecture that when this quantity is extensive, the entanglement entropy generically exhibits volume-law scaling, as in Hermitian systems \cite{kaga_2023, Kaneko_2025}; in contrast, when it remains finite, the entropy obeys a log-volume scaling law. Below, we introduce a concrete model in which such a measurement-induced condensation transition takes place and confirm the applicability of our diagnostic tool.

\section{Entanglement phase transition in the dimer chain} 
\label{sec4}

In this section, we first introduce a monitored dimer Hamiltonian and demonstrate that it undergoes an exceptional-point transition, as evidenced by the long-time behavior of the correlation functions. We then extend this construction to a one-dimensional hopping chain of coupled dimers and investigate its entanglement properties for Fock-state initial conditions using Eq. \eqref{SNNmain}. We show that this model exhibits an entanglement transition, with the late-time R\'enyi entropy scaling changing from volume-law to log-volume-law.

\subsection{Exceptional point transition in the single-site dimer model}\label{subsec:onedimer}

\begin{figure}[t!tbp]
    \centering
    \includegraphics[width=0.28\textwidth]{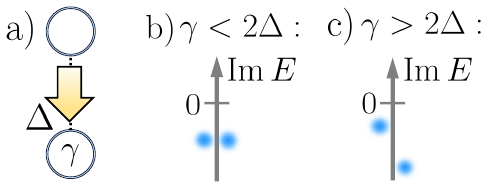}
    \caption{Exceptional point transition of a single monitored dimer (a). When the rate of monitoring $\gamma$ is low, the effective loss on both sites is equal, as evidenced by the equal imaginary parts of the eigenvalues of the Hamiltonian (b). When the monitoring rate $\gamma$ exceeds the coupling $2\Delta$, the sites ``decouple'' with the monitored site characterized by the larger in magnitude value of ${\rm Im} \,  E$ (c). The transition between (a) and (b) occurs at the exceptional point, where the eigenvalues and the eigenvectors of the dimer Hamiltonian merge.}
    \label{fig3}
\end{figure}

\renewcommand{\uparrow}{\psi_0}

First, we consider two isolated sites hybridizing via the hopping $\Delta$, with an imaginary on-site loss term on the bottom site $\gamma$, which describes the no-click measurement \eqref{psitnc}, as sketched in \autoref{fig3}a. The Hamiltonian of this system is 
\begin{equation}
\hat H_{\rm dimer} = \begin{pmatrix}
0 & \Delta \\
\Delta & -i \gamma
\end{pmatrix} \, .
\label{dimham}
\end{equation}
The eigenvalues $E_{1,2} =(- i \gamma \pm \sqrt{4 \Delta^2 - \gamma^2})/2$ become degenerate at the exceptional point $2\Delta = \gamma$. Unlike a typical Hermitian degeneracy, where eigenvectors remain orthogonal and span the Hilbert space, the two eigenvectors of the non-Hermitian Hamiltonian \eqref{dimham} coalesce into a single state. Rather than signaling a topological phase transition, the exceptional point marks a qualitative change in the system’s dynamical behavior: from oscillatory to exponentially purifying \cite{moiseyev2011non,Bergholtz2021}. 

In the case $\gamma<2\Delta$, shown in \autoref{fig3}b, ${\rm Im}\, E_1 = {\rm Im}\, E_2$, and the time decay arising from the non-Hermiticity does not affect observables \eqref{norm} due to the cancellation of the decaying factor by the denominator. As an example, consider a single boson on the top site of \autoref{fig3}, corresponding to the initial density matrix $\hat \rho(0) = \ket{\uparrow} \bra{\uparrow}$, with $\ket{\uparrow} = (1\, , ~ 0)$. According to \eqref{norm}, the expectation value of the  operator $\hat {\cal O} = \ket{\uparrow} \bra{\uparrow}$ at time $t$, determining the probability to remain on the top site, is
\begin{equation}
\braket{\hat{\cal O}}
=  
\frac{\sum_{E,\tilde E} e^{-i(E-\tilde E^*)t} c_E^{\rm L} c_E^{\rm R*} c_{\tilde E}^{\rm L*} c_{\tilde E}^{\rm R}}
{\sum_{E,\tilde E} e^{-i(E-\tilde E^*)t}c_E^{\rm L} c_{\tilde E}^{\rm L*} \braket{\tilde E^{\rm R}|E^{\rm R}} } \, ,
\label{odim}
\end{equation}
where $c_E^{\rm R(L)}= \braket{E^{\rm R(L)}|\uparrow}$. In the considered ${\rm Im}\, E_1 = {\rm Im}\, E_2$ regime, the decay exponent in the denominator cancels that of the numerator, and the behavior of $\braket{\hat {\cal O}}$ is oscillatory, resembling 
a Hermitian system, apart from the subtlety of non-orthonormality $\ket{E^{\rm R}} \neq \ket{E^{\rm L}}$. 

On the other hand, for large monitoring rates $\gamma>2\Delta$, physical observables tend to a time-independent expectation value,
\begin{equation}
\braket{\hat {\cal O}} \to \frac{|\braket{E_1^{\rm R}|\uparrow}|^2}{\braket{E_1^{\rm R}|E_1^{\rm R}}} \, .
\end{equation}
The static nature of the late-time observables in this regime is a manifestation of Zeno localization. As shown below, this effect gives rise to an entanglement phase transition in an extended system composed of coupled monitored dimers.

\subsection{Entanglement phase transition in a dimer chain}\label{subsec:dimerlattice}

\begin{figure}[t!tbp]
    \centering
    \includegraphics[width=0.35\textwidth]{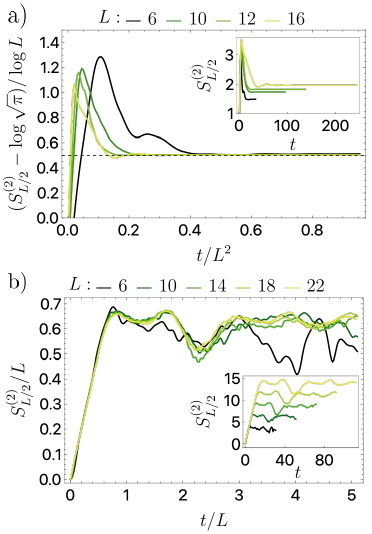}
    \caption{(a), (b): Bipartite R\'enyi entanglement entropy as a function of time for different sizes of the system $L$ corresponding to the $S^{(2)} \sim \log L$ (a) and $S^{(2)} \sim L$ (b) regions (see the phase diagram in \autoref{fig1}b). The figures showcase the collapse of the curves, rescaled by the system length $L$, while the insets show the raw data. (a)  The linear in time entropy build-up phase is followed by the decay of all but one mode. The entanglement follows the predicted $S^{(2)} \simeq \frac{1}{2} \log (\pi L)$ large-$L$ behavior already for very small system sizes (see \cref{appe}). The parameters used are $\Delta = J = J'/2$, $\gamma = 0.3J'$, and $S^{(2)}$ is obtained by the brute-force evaluation of the permanents in \eqref{SNNmain}. (b) After the initial linear growth, the bipartite R\'enyi-2 entropy indefinitely oscillates with time, exhibiting the $S^{(2)} \sim L$ scaling collapse. This behavior resembles that of the systems of free bosons without measurement \cite{kaga_2023, Kaneko_2025}. The parameters used to produce the figure are $J=J'/2$, $\Delta = 5 J'/2$, and $\gamma = 0.2J'$. The results for $L=18$, $22$, are obtained using the random sampling method utilizing the Glynn-type summation \cite{Kaneko_2025}, while smaller sizes are tackled by the brute-force evaluation of the permanents \eqref{SNNmain}.}
    \label{fig4}
\end{figure}

Let us consider a model of many dimers \eqref{dimham} stacked along one direction and add
hoppings $J$ and $J'$ between them, as shown in \autoref{fig1}a. The entire bottom row of sites is subject to the measurement of the number of bosons,  with no-click postselection \eqref{psitnc}. The system is initialized with singly occupied sites of the top sublattice (see \autoref{fig1}a). For this initial condition, the entanglement entropy is determined by a combination of permanents \eqref{SNNmain}.

The effective non-Hermitian Hamiltonian of this model in momentum space is
\begin{equation}
\hat H_{\rm eff} = \begin{pmatrix}
J \cos k & \Delta + J' \cos k \\
\Delta + J' \cos k & - i \gamma + J \cos k
\end{pmatrix} \, ,
\label{dimmodel}
\end{equation}
where the $2 \times 2$ structure arises from the sublattice (top or bottom) degree of freedom.
The eigenvalues of this Hamiltonian are given by,
\begin{equation}
E_{1,2}= J\cos k-i \gamma/2 \pm \sqrt{(\Delta+J'\cos k)^2- \big(\gamma/2\big)^2} \, .
\label{e12dim}
\end{equation}
Similar to the single dimer system \eqref{dimham}, at large inter-chain coupling $\Delta \gg \gamma$, the model \eqref{dimmodel} is in the phase where the time decay is effectively uniform across the sites, and the behavior of the late-time correlators is oscillatory. This gives rise to Region III in the phase diagram in \autoref{fig1}b. In the opposite limit (Regions I and II), the ${\rm Im}\, E(k)$ spectrum is dispersive, and the majority of the modes decay with time, leading to the dynamical Bose-Einstein condensation. Equation \eqref{SNNmain} allows for computing the entanglement entropy in both of these regimes with the initial condition shown in \autoref{fig1}a, as we discuss below.

Let us first consider {\it Region III}, where the argument of the square-root in $\eqref{e12dim}$ is positive for all $k$, and all eigenmodes $k$ having identical imaginary part $\mathrm{Im} \, E(k)=-\gamma/2$. This condition is satisfied for $\Delta-J' \geq \gamma/2$ and highlighted in blue in \autoref{fig1}b. The decaying prefactor arising from the imaginary energy factorizes in both the Green's functions \eqref{gf} and the overlaps \eqref{calgs}. Due to the matching number of $\hat {\cal G}$ contractions in the numerator and denominator in the expression for the entanglement entropy \eqref{SNNmain}, the non-unitary time dependence due to ${\rm Im}\, E(k)$ cancels out. As a result, the behavior of the system resembles that of the Hermitian hopping system of bosons studied in \cite{kaga_2023, Kaneko_2025}, apart from the effects due to the non-orthonormality of eigenstates, which are not able to change the entanglement scaling. In agreement with this earlier work and as confirmed numerically in \autoref{fig4}b by evaluating permanents in \eqref{SNNmain}, after the initial phase of linear in $t$ growth, the bipartite entanglement entropy saturates on the volume-law scaling, $S^{(2)}_{L/2} \sim L$.

In \textit{Regions I and II}, the spectra feature a dispersive imaginary part shown in \autoref{fig1}c. As discussed in \autoref{sscaling}, the late-time entanglement entropy in this regime follows the log-volume scaling due to the formation of the Bose-Einstein condensate in the slowest decaying mode. This scaling is confirmed numerically: evaluating the entropy \eqref{SNNmain} as a function of time, we observe an almost perfect collapse onto logarithmic scaling already at modest system sizes, as shown in \autoref{fig4}a. We note that region I is distinguished from region II by the absence of the dissipation gap, which does not alter the entanglement scaling in our model.

The physical origin of the outlined transition between the volume and log-volume scaling of entropy is the measurement back-reaction on the system. The no-click measurement record corresponds to no particle detections on the monitored sublattice, represented by the bottom row of sites in \autoref{fig1}a. This results in the pile-up of bosons in the mode with the least support on the monitored sites. Indeed, for our effective non-Hermitian Hamiltonian \eqref{dimmodel}, we find
\begin{equation}
{\rm Im}\, E= - \gamma \frac{\braket{E^{\rm R}|\hat P^{\rm B}|E^{\rm R}}}{\braket{E^{\rm R}|E^{\rm R}}}\, ,
\end{equation}
where $\hat P^{\rm B} = {\rm diag}\,\{0,1\}$ is the projector into the monitored sublattice. 

\begin{figure}[t!tbp]
    \centering
    \includegraphics[width=0.35\textwidth]{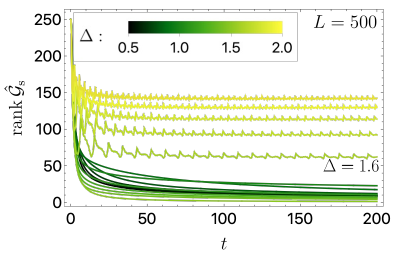}
    \caption{The late-time effective rank \eqref{rk} of the Green's function contraction ${\cal G}_{\rm s}$ entering the numerator in \eqref{SNNmain} is indicative of the measurement-induced phase transition. All curves correspond to $\gamma=J'=2J$. When $\Delta \leq 1.5$, ${\rm rank}\, {\cal G}_{\rm s}$ approaches 1 due to the decay of all but a single eigenmode in the system, while $\Delta > 1.5$ corresponds to the finite $O(L)$ values of the late-time effective rank.}
    \label{fig41}
\end{figure}

As shown previously, the measurement-induced transition between the $S^{(2)}_{L/2} \sim L$ and $S^{(2)}_{L/2} \sim \log L$ scaling regimes is captured by the effective rank of the overlap matrix \eqref{rk}. In \autoref{fig41}, we show this quantity for different values of $\Delta/J'$ at fixed $\gamma/J' = 1$. We observe that in the phase with log-volume entropy scaling, the late-time rank decays to unity, while in the volume-law phase ${\rm rank}\,\hat {\cal G}_{\rm s}(t \to \infty) \gg 1$. Thus, this quantity provides a simple, numerically inexpensive way to diagnose the entanglement transition, which occurs near $\Delta = 1.5$ for the parameter choice in \autoref{fig41}. Lastly, we note that the single-particle correlation matrix $C(x,y,t)$ \eqref{corr1}, often used as a probe of the off-diagonal long-range order accompanying Bose--Einstein condensation, is computationally less efficient to evaluate than the effective rank of the overlap matrix. This is because its computation also involves evaluating matrix permanents.

Interestingly, using a thermal initial state also allows one to effectively diagnose the two entanglement regimes above, which, in this case, manifest as a dynamical purification transition \cite{Gullans2020}, as discussed below and in detail in \cref{appd}. 

\subsection{Dependence of the transition on the initial state}\label{subSec:thermNoclick}

The dynamics of the system \eqref{dimmodel} changes dramatically if the initial on-site Fock state considered above is replaced with a thermal density matrix on sublattice $\rm A$, with the sublattice B initialized in vacuum
\begin{equation}
    \hat{\widetilde{\rho}}_{\mathrm{th}}=e^{-\beta\varepsilon\sum_{x\in \rm A} \hat{b}_x^{\dagger} \hat{b}_x} \, ,
    \label{thmain}
\end{equation}
where $\beta$ and $\varepsilon$ are arbitrary inverse temperature and on-site potentials. For this initial density matrix, the sectors with different numbers of bosons decay at different rates until only the vacuum remains. This occurs due to the back-reaction of the measurement on the state: the absence of boson detections, together with the uncertainty in the initial number of bosons, leads the experimentalist to conclude that the system is empty, resulting in a vacuum steady state.

Despite the decay to vacuum for any combination of parameters in the model \eqref{dimmodel}, the system initialized in a thermal state \eqref{thmain} exhibits a measurement-induced transition in the spirit of Gullans and Huse \cite{Gullans2020pure}. In particular, in the region shaded in green in \autoref{fig1}b, the system approaches the vacuum in a state with area law subsystem R\'enyi entropy, while for parameters that belong to the blue region, the system never purifies before its eventual vacuum decay, exhibiting volume-law scaling of the subsystem R\'enyi entropy on approach to vacuum (see \cref{appd}). Thus, the measurement-induced entanglement transition for the initial Fock state turns into a dynamical purification transition if the initial conditions instead correspond to a thermal density matrix \eqref{thmain}.

\section{Conclusion}

We studied entanglement dynamics of free bosons in the no-click measurement trajectory, described by an effective non-Hermitian Hamiltonian. We obtained analytical expressions for the second R\'enyi entropy for both thermal and on-site Fock initial states. For on-site Fock states, the entropy is given in terms of permanents, which encode the many-body coherence through combinatorial sums of single-particle Green's functions. Although permanents are exponentially hard to evaluate, the key result of this work is that the scaling of the entropy is controlled by a much simpler object: the structure, and in particular the rank, of the matrix entering the permanents.

We further introduced a hopping model with number measurement and no-click post-selection featuring a phase transition from the volume-scaling $S^{(2)}_{L/2} \sim L$ phase to the ``critical'' $S^{(2)}_{L/2} \sim \log L$ phase, as the measurement rate $\gamma$ is increased. The origin of the logarithmic scaling of the entanglement entropy at large measurement rate $\gamma$ is different from that in both monitored free fermion systems and unitary circuits. In free fermions, the $\log L$ scaling is associated with a partial preservation of the Fermi surface within entangled regions, while in circuits $\log L$ arises from the non-unitary critical CFT, obtained via the statistical-mechanical mapping. In contrast, in our system of monitored free bosons, the log-volume scaling arises from Bose-Einstein condensation of all particles into the slowest-decaying mode.

Lastly, we note that the formalism developed here naturally extends to settings with projective measurements ($\hat K_x^{(n)} \sim \ket{n_x}\bra{n_x}$), since the derivative trick used to exponentiate the initial condition can similarly be applied to $\hat K_x^{(n)}$ appearing at a finite time. Thus, we expect that the addition of projective measurements and perturbative interactions will lead to the appearance of similar bosonic permanent constructions, for which our field-theory approach may be fruitful.

\begin{acknowledgments}
\emph{Acknowledgments ---} A.C. thanks Sebastian Diehl and Romain Vasseur for
helpful conversations and acknowledges support from
NSF-2609899. This research was supported in part by grant NSF PHY-2309135 to the Kavli Institute for Theoretical Physics (KITP), where the project was initiated. 
\end{acknowledgments}

\bibliography{litimp}

\pagebreak
\onecolumngrid
\newpage
\appendix
\crefalias{section}{appendix}
\begin{center}
\textbf{\large Supplementary information for\\ ``Theory of post-selected entanglement transitions in monitored bosons"}
\\[6pt]
Ilia Komissarov, Emanuele Dalla Torre, Ahana Chakraborty
\end{center}

\renewcommand{\thefigure}{\arabic{figure}}
\renewcommand{\figurename}{Supplementary Figure}
\renewcommand{\tablename}{Supplementary Table}
\setcounter{figure}{0}

In this Supplementary Information, we first review some important spectral properties of non-Hermitian Hamiltonians. Then, we review the expression for the second R\'enyi entanglement entropy via the contraction of two Wigner characteristic functions. After this, we apply this formalism to compute the entanglement entropy as a function of time in non-Hermitian systems with thermal initial conditions. We further generalize our treatment towards the systems of bosons initialized in on-site Fock states, and obtain an analytic expression for the entanglement entropy in two cases: one corresponding to $N$ bosons initialized on a single site, and the other describing one boson initialized on each of $N$ sites. We also consider the entanglement entropy in a non-Hermitian bosonic system initialized in a mixture of thermal and on-site Fock states.

\section{Non-Hermitian quantum mechanics}

\label{appa}

In this Appendix, we discuss the main differences between Hermitian and non-Hermitian systems with respect to their eigenstates and spectra. We further discuss the differences between left- and right-eigenstates and the physical definition of density matrices in non-Hermitian systems. For a broader review, see \cite{Ashida_2020}

Eigenvalues of a general non-Hermitian Hamiltonian are complex-valued. A notable exception occurs in systems invariant under the combined action of parity $\mathcal P$ and time-reversal $\mathcal T$: in this case, the on-site losses modeled by the negative imaginary on-site potentials $- i \gamma$, if present, are compensated by on-site gains \cite{LeGal2023}. In this work, we primarily focus on systems subject to continuous measurement of the local particle number in the no-click limit, which are described using on-site loss terms only. We thus inevitably have to consider systems that are intrinsically non-$\mathcal{PT}$-invariant.

Apart from the presence of complex eigenvalues, which lead to non-unitary time evolution, an additional distinction from Hermitian quantum mechanics arises from the fact that the eigenvectors of generic non-Hermitian Hamiltonians do not form an orthonormal basis. Indeed, from 
\begin{equation}
\hat H \ket{E^{\rm R}} = E \ket{E^{\rm R}}  \, , \qquad \hat H \ket{\tilde E^{\rm R}} = \tilde E \ket{\tilde E^{\rm R}}\, , \qquad E \neq \tilde E \, ,
\label{HR}
\end{equation}
one cannot, in general, prove $\braket{E^{\rm R}|\tilde E^{\rm R}}$. One exception is the Hermitian Hamiltonian with added uniform on-site imaginary potentials $\hat H_{\rm eff} =  \hat H_{\rm herm} - i \gamma \hat I$: this situation corresponds to the no-click measurement applied uniformly across all sites, as considered below (see \cref{appc}). The superscript ``R'' on the eigenvectors in \eqref{HR} stands for ``right'': left eigenvectors are defined according to
\begin{equation}
\hat H^{\dagger} \ket{E^{\rm L}} = E^{*} \ket{E^{\rm L}} \, , \qquad \bra{E^{\rm L}} \hat H = E \bra{E^{\rm L}} \, .
\end{equation}
While left and right sets of eigenvectors are typically separately non-orthonormal, one can usually introduce a doubled (bi-orthonormal) basis consisting of both right and left eigenvectors, which satisfies
\begin{equation}
\braket{E^{\rm L}| \hat H |\tilde E^{\rm R}} = E \delta_{E \tilde E} \, , \qquad
\braket{E^{\rm L}|\tilde E^{\rm R}} = \delta_{E \tilde E} \, , \qquad
\sum_{E} \ket{E^{\rm R}} \bra{E^{\rm L}} =  \hat I \, ,
\label{psirl}
\end{equation}
where $\hat I$ is the unit matrix. The construction above may still fail at certain fine-tuned parameter values in non-Hermitian Hamiltonians, known as the \textit{exceptional points} \cite{Ding_2022}.

In analogy with the introduced right and left Hamiltonian eigenstates $\ket{E^{\rm R,L}}$, it is convenient to consider generic right and left state vectors, distinguished by their time evolution,
\begin{equation}
\ket{\psi^{\rm R}(t)} = e^{- i \hat H t} \ket{\psi^{\rm R}(0)} \, , \qquad
\ket{\psi^{\rm L}(t)} = e^{- i \hat H^\dagger t} \ket{\psi^{\rm L}(0)} \, ,
\end{equation}
where we assumed the time independence of the non-Hermitian Hamiltonian. Density matrices defined from the state vectors in mixed left–right bases are often more convenient to evolve with time. As an illustrative example, consider the trace of an arbitrary pure-state ``density matrix'' defined in the right–left basis,
\begin{equation}
{\rm Tr}\!\left[\hat{\rho}^{\rm RL}(t)\right]
= {\rm Tr}\!\left[\ket{\psi^{\rm R}(t)} \bra{\psi^{\rm L}(t)}\right]
= {\rm Tr}\!\left[e^{- i \hat H t} \ket{\psi^{\rm R}} \bra{\psi^{\rm L}} e^{i \hat H t}\right]
= \sum_{E} \braket{E^{\rm L}|\psi^{\rm R}} \braket{\psi^{\rm L}|E^{\rm R}} \, .
\label{rrlt}
\end{equation}
The expression above is time-independent and involves only a single spectral summation, which makes objects such as $\hat \rho^{\rm RL}$ technically convenient for studying entanglement in non-Hermitian systems. Nonetheless, $\hat \rho^{\rm RL}$ is manifestly non-Hermitian and therefore cannot represent a physical state of the system. Moreover, Eq.~\eqref{rrlt} does not capture the non-unitary probability decay in monitored systems dictated by the Born rule manifest in the non-Hermiticity of $\hat H$.

For these reasons, we restrict our attention to density matrices defined in the RR basis, which obey the expected quantum-mechanical time evolution,
\begin{equation}
{\rm Tr}\!\left[\hat{\rho}^{\rm RR}(t)\right]
= {\rm Tr}\!\left[\ket{\psi^{\rm R}(t)} \bra{\psi^{\rm R}(t)}\right]
= {\rm Tr}\!\left[e^{- i \hat H t} \ket{\psi^{\rm R}} \bra{\psi^{\rm R}} e^{i \hat H^\dagger t}\right]
= \sum_{E, \tilde E}
\braket{E^{\rm L}|\psi^{\rm R}}
\braket{\psi^{\rm R}|\tilde E^{\rm L}}
\braket{\tilde E^{\rm R}|E^{\rm R}}
e^{- i (E - \tilde E^*) t} \, .
\label{rrrt}
\end{equation}

Note the double summation over energy eigenvalues in the expression above. It originates from the fact that a general non-Hermitian Hamiltonian can be diagonalized only in the biorthogonal basis \eqref{psirl}. In contrast, Hermitian finite-time density matrices must be constructed solely from right eigenvectors. In addition, due to complex eigenvalues ${\rm Im}\, E < 0$, the terms in \eqref{rrrt} decay in time. These features require modifications to the expressions used to compute the entanglement properties of Hermitian systems, as we elaborate below.

\section{R\'enyi entanglement entropy from Wigner characteristic function in a general non-Hermitian system}

\label{appb}

The Wigner characteristic function (WCF) $\chi^{\rm W}(t,\{\xi_x\})$ provides tomographic information about entanglement in systems of bosons \cite{CG,WCF2,WCF3}, with $\{\xi_x\}=\{\xi_1,\xi_1^*,\xi_2,\xi_2^*,\ldots\}$ playing a role of classical phase-space variables. In Hermitian systems, the WCF is defined as
\begin{equation}
\chi^{\rm W}(t,\{\xi_x\})={\rm Tr}\,[\hat\rho(t)\hat D(\{\xi_x\})]\,, \qquad
\hat D(\{\xi_x\})\equiv
e^{\sum_x(\xi_x \hat b_x^{\dagger}-\xi_x^*\hat b_x)} \, ,
\label{wcfgen}
\end{equation}
where $\hat D(\{\xi_x\})$ are referred to as the displacement operators, and
$\hat b_x$ and $\hat b_x^\dagger$ are the annihilation and creation operators of bosons
on site $x$. The Wigner characteristic function provides a convenient way to compute the R\'enyi entanglement entropy in systems of bosons, bypassing the need for complicated time-evolution contour geometries arising from traces of density matrices over subsystems \cite{Chakraborty_2021}.

We define the RR Wigner characteristic function by replacing $\hat\rho(t)$ in
Eq.~\eqref{wcfgen} with the RR density matrix,
\begin{equation}
\chi^{\rm W, \rm RR}(t,\{\xi_x\})={\rm Tr}\,[\hat\rho^{\rm RR}(t)\hat D(\{\xi_x\})] \, .
\label{wcfdef}
\end{equation}
The second R\'enyi entropy of a non-Hermitian system is expressed as \cite{CG}
$\hat\rho^{\rm RR}_{\rm s}={\rm Tr}_{\overline{\rm s}}[\hat\rho^{\rm RR}]$ as
\begin{equation}
\begin{aligned}
e^{-\tilde S^{(2)}_{\rm s}(t)}
&={\rm Tr}\!\left[(\hat\rho^{\rm RR}_{\rm s})^2\right] \\
&={\rm Tr}\!\left[
\int\!\left\{\prod_{y\in{\rm s}}\frac{d\xi_y\,d\xi_y^*}{\pi}\right\}
{\rm Tr}\!\left[\hat D(\{\xi_x\})\hat\rho^{\rm RR}_{\rm s}\right]
\hat D(\{-\xi_x\})\hat\rho^{\rm RR}_{\rm s}
\right] \\
&=\int\!\left\{\prod_{y\in{\rm s}}\frac{d\xi_y\,d\xi_y^*}{\pi}\right\}
\chi^{\rm W,\,\rm RR}(t,\{-\xi_x\})\,\chi^{\rm W,\,\rm RR}(t,\{\xi_x\}) \, ,
\end{aligned}
\label{tildeS}
\end{equation}
where $\overline{\rm s}$ denotes the complement of ${\rm s}$. Here we used the fact that the operators $\hat D(\{\xi_{x\in{\rm s}}\})$
form an orthonormal basis in the Hilbert space of operators with support in ${\rm s}$,
\begin{equation}
{\rm Tr}\!\left[
\hat D(\{\xi_{x\in{\rm s}}, \xi_{x\in{\overline {\rm s}}} =0 \})\hat D(\{-\gamma_{x\in{\rm s}}, -\gamma_{x\in \overline {\rm s}} = 0\})
\right]
=\prod_{x\in{\rm s}}\pi\,
\delta(\xi_x-\gamma_x)\,
\delta(\xi_x^*-\gamma_x^*) \, .
\end{equation}
The expression \eqref{tildeS} is not final, as it does not account for the non-unitary nature of the time evolution. Density matrices of non-Hermitian systems generally decay with time, violating the normalization condition ${\rm Tr}\, \hat \rho^{\rm RR}(t) = 1$. For example, if a constant loss term is introduced at every site $\hat H_{\rm eff} = \hat H -i \gamma \sum_x \hat b^\dagger_x \hat b_x$, the initial Fock state occupied by $N$ bosons decays in time as
\begin{equation}
\hat \rho^{\rm RR}(t) = e^{- 2 \gamma N t} \hat \rho^{\rm RR}(0)\, .
\end{equation}
Thus, every density matrix entering an observable should be normalized by ${\rm Tr}\, \hat \rho^{\rm RR}(t)$ to restore the normalization of probability. Since the second R\'enyi entropy is defined using two density matrices, the normalized expression is
\begin{equation}
e^{-S^{(2)}_{\rm s}(t)}
=
\frac{\displaystyle
\int\!\left\{\prod_{y\in{\rm s}}\frac{d\xi_y\,d\xi_y^*}{\pi}\right\}
\chi^{\rm W,\,\rm RR}(t,\{-\xi_x\})\,\chi^{\rm W,\,\rm RR}(t,\{\xi_x\})
}{
\left({\rm Tr}\,[\hat\rho^{\rm RR}(t)]\right)^2
}
\, .
\label{Sfin}
\end{equation}

This is the main expression that we use in this work to compute the entanglement entropy of bosons. Since we mainly operate with quantities defined in the right-right basis, in the following we drop the index R everywhere apart from the energy eigenstates $\ket{E^{\rm L}}$, $\ket{E^{\rm R}}$ appearing in eigenmode expansions. Note that the only input required to calculate $S^{(2)}_{\rm s}(t)$ is the Wigner characteristic function, since
${\rm Tr}\,[\hat\rho(t)]=\chi^{\rm W}(t,\{\xi_x=0\})$.

To compute $\chi^{\rm W}(t,\{\xi\})$, it is useful to represent the trace in Eq.~\eqref{wcfdef} as a double-branch Keldysh path integral (see \autoref{fig2}b) over on-site coherent states $\ket{\{\phi_{\tau}(1),\ldots,\phi_{\tau}(L)\}}$ defined at a time slice $\tau$ and satisfy
\begin{equation}
\hat b_i\ket{\{\phi_{\tau}(1),\ldots,\phi_{\tau}(L)\}}=\phi_{\tau}(i)\ket{\{\phi_{\tau}(1),\ldots,\phi_{\tau}(L)\}} \, ,
\end{equation}
where we assume a discrete system in the real space with $L$ sites. 

As shown in the main text, \eqref{wcfdef} can be written as a path integral over the coherent states, which can be evaluated analytically in Gaussian systems of bosons, which yields
\begin{equation}
\chi^{\rm W}(t,\{\xi_x\})
={\rm Tr}\,[\hat\rho(t)]
\exp\!\left[-\frac{i}{2}\sum_{x,y}\xi_x^*
G_{\rm K}(x,y,t,t,t)\xi_y\right] \, .
\label{chiRRfin}
\end{equation}
We note that Fock initial conditions of bosons violate the assumption of Gaussianity, and the expression above no longer applies. In this case, the initial density matrix must be suitably exponentiated, and additional
manipulations of Eq.~\eqref{chiRRfin} are required, as discussed in Appendix~\ref{appe}.

To summarize, Eqs.~\eqref{Sfin} and \eqref{chiRRfin} provide a general framework for
computing the second R\'enyi entanglement entropy in Gaussian non-Hermitian bosonic systems.
In the following sections, we illustrate the applications of this formalism in a sequence of models of
increasing complexity. We first consider unitarily diagonalizable systems with Gaussian (thermal) initial conditions, followed by non-Hermitian systems with thermal initial states, where no orthonormal set of right eigenvectors exists. We then address non-Hermitian systems with bosons initialized in Fock states. 

\section{Unitarily diagonalizable systems with thermal initial conditions} 

\label{appc}

As a simplest example, we consider the class of Hamiltonians for which the set of right and left eigenstates coincide, $\ket{E^{\rm L}} = \ket{E^{\rm R}} = \ket{E}$. This implies that the Hamiltonian can be diagonalized by a unitary transformation $\hat H \to \hat U^\dagger \hat H \hat U$, where the columns of $\hat U$ are $\ket{E}$. This condition does not imply that the considered model is Hermitian, which is exemplified by the following hopping Hamiltonian:
\begin{equation}
\hat H = J \sum_x \hat b^\dagger_x \hat b_{x+1} + {\rm h.c.} - i \gamma \sum_x \hat b^\dagger_x \hat b_x \, .
\label{hamun}
\end{equation}
This model describes bosons hopping on a 1D chain, with the number of bosons on each site continuously monitored at a constant rate $\gamma$, with a post-selection of a trajectory in which no boson losses occurred \cite{Biella_2021, Turkeshi_2021}. In this model, the amplitude of a state with the total of $N$ bosons decays in time as $e^{- \gamma N t}$, since
\begin{equation}
\frac{d}{dt} \ket{N(t)} = - i \hat H \ket{N(t)} = - \gamma N \ket{N(t)} + \ldots \, ,
\label{ndecay}
\end{equation}
where the $\ldots$ include the unitary evolution part arising from the hopping terms in \eqref{hamun}.

Left and right eigenstates of $\hat H$ defined in \eqref{hamun} are regular plain waves $\ket{k}$, while the eigenvalues contain a constant imaginary part
\begin{equation}
E(k) \equiv E'(k) + i E''\, , \qquad E'(k) = 2 J \cos k\, , \qquad E'' = - \gamma \, .
\end{equation}
We are interested in the dynamics of entanglement in the model \eqref{hamun} with thermal initial conditions with inverse temperature $\beta$ and chemical potential $\mu$
\begin{equation}
\hat \rho_{\rm th} = e^{-\beta (\hat H_{\rm Herm} - \mu \hat N )}\, ,
\label{rhoth}
\end{equation}
where $\hat H_{\rm Herm} = (\hat H + \hat H^\dagger)/2$ is the Hamiltonian of the system in the absence of measurement and $\hat N = \sum_x \hat b^\dagger_x \hat b_x$. Using $\hat H_{\rm Herm}$ ensures the Hermiticity of the density matrix. This setup corresponds to the system initially prepared in thermal equilibrium in the absence of measurement with $\gamma$ subsequently turned on. We ignore the normalization factor in \eqref{rhoth}, as it does not affect the normalized entanglement entropy \eqref{Sfin}. Since $\hat H$ and $\hat H^\dagger$ are both unitarily diagonalizable in the same (momentum) basis, it is straightforward to write down expressions for expectation values of operators in terms of the discretized action in the momentum coherent-state basis
\begin{equation}
\hat b_k \ket{\phi_{\tau}({k_1}), \ldots, \phi_{\tau}({k_L})} = \phi_\tau(k) \ket{\phi_{\tau}({k_1}), \ldots, \phi_{\tau}({k_L})} \, ,
\end{equation}
where $k_x$ are the allowed momenta in the discrete system. We find
\begin{equation}
{\rm Tr}\,[\hat \rho(t)] = {\rm Tr}\,[e^{i \hat H^\dagger t}\hat \rho_{\rm th}e^{-i \hat H t}]
=\int D\phi^+ D\phi^- \,
e^{i S(t)} \, ,
\label{parf}
\end{equation}
where
\begin{equation}
i S(t)
=  \sum_{\tau_1,\tau_2}^{2T} \sum_{k} \delta_t^2 \,
\phi_{\tau_1}^{*}(k)\, i G_{\tau_1,\tau_2,T}^{-1}(k)\, \phi_{\tau_2}(k) \, ,
\end{equation}
and $\phi_{\tau} = \phi^+_\tau$ for $1 \leq \tau \leq T$, $\phi_{\tau} = \phi^-_\tau$ for $T+1 \leq \tau \leq 2T$, with the inverse Green's function matrix having the following structure, shown for $T=3$ as an example:
\begin{equation}
[i G^{-1}(k)]_{T=3} = 
\left[
\begin{array}{ccc|ccc}
-1 & 0 & 0 & 0 & 0 & \rho(k) \\
H^-(k) & -1 & 0 & 0 & 0 & 0 \\
0 & H^-(k) & -1 & 0 & 0 & 0 \\ \hline
0 & 0 & 1 & -1 & 0 & 0 \\
0 & 0 & 0 & H^+(k) & -1 & 0 \\
0 & 0 & 0 & 0 & H^+(k) & -1
\end{array}
\right] \, ,
\end{equation}
with $H^{\mp}(k) \equiv 1 \mp i E'(k)\delta_t - E'' \delta_t$, $\rho(k) = e^{-\beta[E'(k) - \mu]}$, and $\delta_t \cdot T = t$. Computing the determinant, we establish for the model \eqref{hamun} in the continuum limit $\delta_t \to 0$, $T \to \infty$, $\delta_t \cdot T = t$:
\begin{equation}
{\rm Tr} [\hat \rho(t)] = \prod_k \frac{1}{\det i G^{-1}(k,t)}
= \prod_k \frac{1}{1 - e^{-\beta [E'(k)-\mu] - 2\gamma t}} \, .
\label{trrtherm}
\end{equation}
At $t=0$, this expression reproduces the partition function of an ensemble of independent oscillators with energies $E(k)$ \cite{kam}, while ${\rm Tr}\,\hat \rho(t) \to 1$ at $t \to \infty$, which signifies the decay of the components of the mixture with non-zero occupation number in momentum space, with only the vacuum remaining at large times, $\hat \rho(t) \to \ket{0}\bra{0}$.

To study the entanglement in this model, we determine the Keldysh Green's function, which depends on the time variables $\tau_1$ and $\tau_2$ running along the contour and the time corresponding to the tip of the contour $t = \delta_t \cdot T$:
\begin{equation}
G^{\rm K}_{\tau_1,\tau_2,T}(k) =
\frac{\rho(k) (H^+)^{\tau_2-1} (H^-)^{\tau_1-1}
+ (H^+ H^-)^{T-1} (H^+)^{1-\tau_1} (H^-)^{1-\tau_2}}
{\det i G^{-1}(k,t)} \, .
\end{equation}
In the continuum limit, we find for a general unitarily diagonalizable non-Hermitian system with the initial condition \eqref{rhoth}
\begin{equation}
G^{\rm K}(k,t_1,t_2,t)
= \frac{e^{(t_1+t_2-2t)E'' } + e^{-\beta [E'-\mu] + 2 (t_1+t_2) E''}}
{1 - e^{-\beta [E'-\mu] + 2 t E''}} \, .
\label{GKuni}
\end{equation}
Note the explicit dependence on the time corresponding to the tip of the Keldysh contour $t$ (see \autoref{fig2}b). Using \eqref{GKuni}, observables in non-Hermitian systems can be determined, such as the occupation of the state with momentum $k$ at time $t$
\begin{equation}
\braket{\hat n(k,t)} = {\rm Tr}\,[\hat b^\dagger_k \hat b_k \hat \rho_{\rm th}(t)]
= \frac{1}{2}\bigl(G_{\rm K}(k,t,t,t)-1\bigr)
= \frac{e^{-\beta [E'-\mu]+ 2 t E''}}{1 - e^{-\beta [E'-\mu] + 2 t E''}}
= e^{-\beta [E'-\mu] + 2 t E''} + e^{-2\beta [E'-\mu] + 4 t E''} + \ldots \, ,
\label{nk}
\end{equation}
where $e^{-2 N \beta [E'(k)-\mu]}$ is the Boltzmann factor describing the probability of populating the state with energy $E'(k)$ by $N$ bosons, and the factor $e^{2 N E''(k) t} = e^{-2 N \gamma t}$ in our model \eqref{hamun}, where $E''(k) = -\gamma$, representing the time decay of the state occupied by $N$ bosons (see \eqref{ndecay}).

The knowledge of the Keldysh Green's function allows us to obtain the second R\'enyi entanglement entropy by evaluating the integral over $\{\xi_x\}$ in the expression \eqref{Sfin}:
\begin{equation}
S^{(2)}_{\rm s}(t)
= {\rm Tr} \log \left[
\sum_{k} \psi^*_k(x)\, i G^{\rm K}(k,t,t,t)\, \psi_k(y)
\right]_{x,y \in \rm s} \, , \qquad 
i G^{\rm K}(k,t,t,t)
= \frac{1 + e^{-\beta [E'-\mu] + 2 t E''}}
{1 - e^{-\beta [E'-\mu] + 2 t E''}} \, .
\end{equation}
At $t=0$, this expression matches the second R\'enyi entropy of free bosons at finite temperature, with the bipartite entropy scaling linearly with the system size $L$ \cite{ACN}. At $t \to \infty$, in our model \eqref{hamun}, the initial thermal state decays into vacuum with $S^{(2)} = 0$; on approach to the pure state, the bipartite entropy exhibits volume-law scaling:
\begin{equation}
S^{(2)}_{L/2}(t \to \infty)
\simeq {\rm Tr} \left[ \sum_k
\psi^*_k(x)\, \bigl(2 e^{-\beta [E'-\mu] + 2 t E''}\bigr)\, \psi_k(y) \right]_{x,y \in \rm s}
= e^{-2\gamma t} \sum_k e^{-\beta [E'(k)-\mu]}
=2 I_0(2\beta J)\, e^{-2\gamma t} e^{ \beta \mu} L \, ,
\end{equation}
where $I_0(x)$ is the modified Bessel function of the first kind. The scaling with $L$ is an artifact of the absence of dispersion in the imaginary spectrum ($E''(k) = - \gamma$) in the considered model \eqref{hamun}: as we show below, once the dispersion in $E''(k)$ is introduced, the initial thermal density matrix purifies, and the late-time entropy follows the area law scaling. To establish this result, we extend our formalism to non-unitarily diagonalizable non-Hermitian Hamiltonians.

\section{LR-diagonalizable systems with thermal initial conditions} 

\label{appd}

General non-Hermitian Hamiltonians cannot be diagonalized by a unitary transformation, and instead a transformation $\hat H \to \hat L^\dagger \hat H \hat R$ needs to be utilized, where the matrices $\hat R$ and $\hat L$ are composed from the right and left eigenvectors of the Hamiltonian $\hat H$ \eqref{psirl}. The sets of right and left eigenvectors of the Hamiltonian in such systems are no longer separately orthonormalized, as discussed in \cref{appa}. A convenient initial state to consider in this case is the thermal density matrix
\begin{equation}
\hat{\widetilde \rho}_{\rm th} = e^{-\sum_x \beta_x (\varepsilon_x - \mu) \hat b^\dagger_x \hat b_x}\, ,
\label{trhth}
\end{equation}
corresponding to the system in which each site is initialized with its own inverse temperature $\beta_x$. The on-site potentials $\varepsilon_x$ can also vary from site to site. Here, we address the dynamics of entanglement in such a system undergoing general non-unitary time evolution. The density matrix $\hat{\widetilde \rho}_{\rm th}$ can be exponentiated in the on-site coherent state basis, contributing to the top-right block in the Keldysh action, which couples the bosonic fields at the forward and backward branches at the initial time $\tau = 0$
\begin{equation}
iS^{+-} = \sum_{\tau_1,\tau_2} \sum_{x,y} \delta_t^2 \phi^{+*}_{\tau_1}(x) i\Delta_{\tau_1,\tau_2}^{+-}(x,y) \phi^-_{\tau_2}(y)  \, , \qquad \Delta_{\tau_1,\tau_2}^{+-}(x,y) =-i e^{-\beta_x(\varepsilon_x - \mu)} \delta_{x,y} \delta_{\tau_1,0}\delta_{\tau_2,0} \equiv -i \Delta(x,y) \delta_{\tau_10}\delta_{\tau_20} \, .
\label{spmtherm}
\end{equation}

The necessary ingredients to determine the entanglement entropy \eqref{Sfin} are the norm of the density matrix and the equal-time Keldysh Green's function evaluated at the time corresponding to the tip of the contour $t$. The former quantity is obtained by the direct evaluation of the determinant of the Keldysh action
\begin{equation}
  {\rm Tr}\, \hat \rho(t) = {\rm det}^{-1}(i \hat G^{-1}) \, .
\label{trrdet}
\end{equation}
We split the action under the determinant into the action with the vacuum initial condition and the contribution from $\hat{\tilde \rho}_{\rm th}$
\begin{equation}
i \hat G^{-1} = i \hat G_{\rm vac}^{-1} + i \hat \Delta^{+-}\, ,
\label{gm1}
\end{equation}
Plugging this into \eqref{trrdet}, and using $\det(i \hat G_{\rm vac}) = 1$, we find
\begin{equation}
{\rm Tr}\, \hat \rho(t) = {\rm det}^{-1} (\hat I- \hat {\cal G}_{\rm s + \overline s}\hat \Delta) \, ,
\label{trrsitetemp}
\end{equation}
where 
\begin{equation}
{\cal G}_{\rm s + \overline s}(x,y) \equiv iG_{\rm vac}^{-+}(x,y,0,0,t) = \braket{0|\hat b_x e^{i \hat H^\dagger t} e^{-i \hat H t} \hat b^\dagger_{y} |0} = \sum_{E_1,E_2} \psi^{\rm L}_{E_1}(x) \psi^{\rm L*}_{E_2}(y) \braket{E_1^{\rm R}|E_2^{\rm R}} e^{i (E_1^*-E_2) t} \, .
\label{Gt}
\end{equation}

As a check, in a system described by a Hermitian Hamiltonian, the left and right eigenvectors coincide, and the time dependence drops out with $G_{\rm vac}^{-+}(x,y,0,0,t) = - i \delta_{xy}$, with \eqref{trrsitetemp} reducing to the expected time-independent expression
\begin{equation}
{\rm Tr}\, \hat \rho(t) \to \prod_x \frac{1}{1-e^{- \beta_x(\varepsilon_x - \mu)}}\, .
\end{equation}
In the example of the unitarily-diagonalizable non-Hermitian system studied above (see \cref{appc}), $G_{\rm vac}^{-+}(x,y,0,0,t) = - i \delta_{x,y} e^{-2 \gamma t}$, and the real-space analog of \eqref{trrtherm} follows
\begin{equation}
{\rm Tr}\, \hat \rho(t) \to \prod_x \frac{1}{1-e^{- \beta_x(\varepsilon_x - \mu) - 2 \gamma t}}\, ,
\end{equation}
which approaches unity if the elements of the mixture with nonzero boson numbers undergo no-click decay (see \cref{appc}).

The next step is to determine the equal-time Keldysh Green's function
\begin{equation}
G^{\rm K}(x,y,t,t,t) = G^{+-}(x,y,t,t,t)+ G^{-+}(x,y,t,t,t)\, .
\label{kpmmp}
\end{equation}
Each of the terms on the right-hand side is obtained from a Dyson-like expansion following from the decomposition \eqref{gm1}: for instance,
\begin{equation}
\begin{aligned}
G^{-+}(x,y,t,t,t) &= G^{-+}_{\rm vac}(x,y,t,t,t) +i \sum_{z_1} G^{-+}_{\rm vac}(x,z_1,t,0,t) e^{-\beta(\varepsilon_{z_1} - \mu)}  G^{-+}_{\rm vac}(z_1,y,0,t,t) \\
&-i \sum_{z_1, z_2, z_3} G^{-+}_{\rm vac}(x,z_1,t,0,t) e^{-\beta_{z_1}(\varepsilon_{z_1} - \mu)}  G^{-+}_{\rm vac}(z_1,z_2,0,0,t) \Delta(z_2,z_3) G^{-+}_{\rm vac}(z_3,y,0,t,t) + \ldots \\
& = - i \delta_{x,y} - i \sum_{z_1, z_2} G^{>}(x,z_1) e^{-\beta_{z_1}(\varepsilon_{z_1} - \mu)} (\hat I-\hat {\cal G}_{\rm s + \overline s} \hat \Delta)^{-1}(z_1, z_2) G^{>\dagger}(z_2,y)\, ,
\end{aligned}
\end{equation}
where we introduced two more ``greater'' vacuum Green's functions
\begin{equation}
\begin{aligned}
G^{>}(x,y) & \equiv G^{-+}_{\rm vac}(x,y,t,0,t) =\braket{0| e^{i \hat H^\dagger t} \hat b_x e^{-i \hat H t} \hat b^\dagger_{y} |0} =\sum_E \psi^{\rm R}_E(x) \psi^{\rm L*}_E(y) e^{- i E t}
 \, , \\
 G^{> \dagger}(x,y) & \equiv G^{-+}_{\rm vac}(x,y,0,t,t) =\braket{0|\hat b_x e^{i \hat H^\dagger t} \hat b^\dagger_{y} e^{-i \hat H t}  |0} = \sum_E \psi^{\rm L}_E(x) \psi^{\rm R*}_E(y)  e^{i E^* t}\, ,
\end{aligned}
\label{Gg}
\end{equation}
where ${\cal G}_{\rm s + \overline s} = G^{>\dagger} G^{>} $ (see \eqref{Gt}). Obtaining $G^{+-}(x,y,t,t,t)$ in a similar way, and plugging into \eqref{kpmmp}, we obtain the Keldysh Green's function in terms of the vacuum Green's functions \eqref{Gt}, \eqref{Gg}
\begin{equation}
G^{\rm K}(x,y,t,t,t) = - i \delta_{x,y} - 2 i \sum_{z_1, z_2} G^{>}(x,z_1) e^{-\beta(\varepsilon_{z_1} - \mu)} (\hat I-\hat {\cal G}_{\rm s + \overline s} \hat \Delta)^{-1}_{z_1, z_2} G^{>\dagger}(z_2,y) \, .
\label{gkons}
\end{equation}
As a final step, we plug \eqref{gkons}, \eqref{trrdet} into the expression for the WCF \eqref{chiRRfin}, which we use to determine the entanglement entropy \eqref{Sfin}. Performing the gaussian integral over the variables $\{\xi_x\}$ yields a bosonic determinant, and the final expression
\begin{equation}
S_{\rm s}^{(2)} = \mathop{\rm Tr}\log \left[i G^{\rm K}(x,y,t,t,t)\right]_{x,y \in {\rm s}} \, .
\label{sg}
\end{equation}

\begin{figure*}[t!tbp]
    \centering
    \includegraphics[width=1\textwidth]{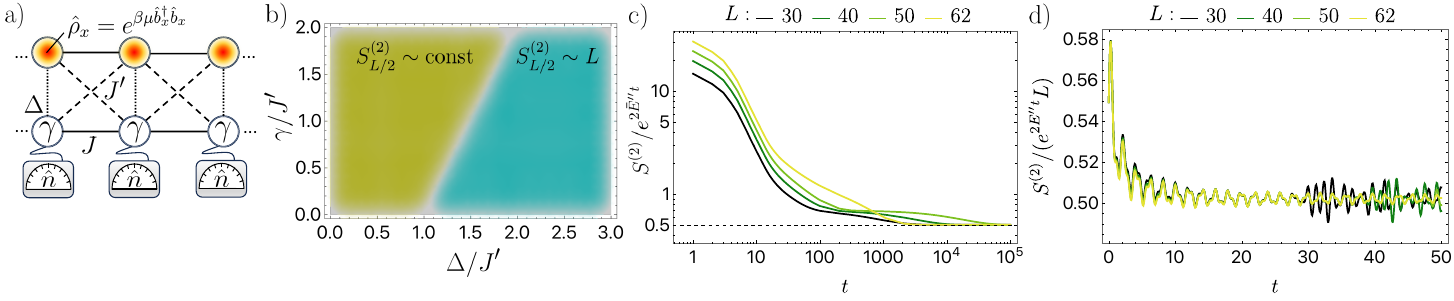}
    \caption{a) We now consider the cross-stitch lattice system (same as in the main text) with initial conditions corresponding to the on-site classical thermal mixture. b) The phase diagram now features two regions: the area law phase, shaded with yellow, and the volume law phase, shaded with blue. c) Dependence of the bipartite second R\'enyi entropy on time for different sizes of the system $L$ does not scale with the size of the system in $t \to \infty$ limit. Parameters used are $\gamma=0.3$, $\Delta=0.6$. d) Volume law scaling of the second R\'enyi entropy. Parameters used are $\gamma=0.3$, $\Delta=2.6$. In both (a) and (c), we use $J'=2J=1$, $e^{\beta \mu} = 1/2$.}
    \label{appfig2}
\end{figure*}

\subsection{Example}

As one application, we consider the dynamics of entanglement in the hopping model of bosons introduced in the main text (see Fig.\ref{fig1}(a)):
\begin{equation}
\hat H_{\rm eff} = \begin{pmatrix}
J \cos k & \Delta + J' \cos k \\
\Delta + J' \cos k & - i \gamma + J \cos k
\end{pmatrix} 
\label{dimapp}
\end{equation}
with the initial conditions corresponding to the quench from the uniform temperature on the isolated sites \eqref{trhth}. The late-time entanglement entropy in this model, initialized with the Fock state, either scales with $\log L$ (phases I,II) or with $L$ (phase III), depending on the parameters of the model (see the main text for the discussion). Here, we show that when initialized in the thermal state \eqref{trhth}, the density matrix eventually purifies in phases I and II, while remaining mixed in phase III. The model \eqref{dimapp}, initialized in a mixed state, thus exhibits a dynamical purification phase transition \cite{Gullans2020pure}, similarly to a non-Hermitian spin model studied in \cite{Gopalakrishnan2021}. In \eqref{trhth}, we set $\beta_x \equiv \beta$ for all the sites at the A sublattice, and $\beta_x \to \infty$ for the B sublattice, although this choice does not affect the scaling of the entanglement entropy at $t \to \infty$. The model \eqref{dimapp} then features two regimes:\\

\underline{\textit{Phase I,II: $E''(k)$ dispersive}} (purifying):

To obtain the leading at $t \to \infty$ term in entanglement entropy, we keep the slowest decaying mode $\chi^{\rm R}(x) =  \psi^{\rm R}_{\overline E} (x)$, $\overline E = \mathop{\rm max}\limits_{E''} E$, in the spectral summations defining the vacuum Green's functions in \eqref{gkons}. The Keldysh Green's function \eqref{gkons} then reduces to
\begin{equation}
\left.i G^{\rm K}(x,y,t,t,t)\right|_{t \to \infty} \simeq \delta_{x,y} + 2 e^{2 \overline E'' t} e^{\beta \mu} \chi^{\rm R}(x) \chi^{\rm R*}(y) \sum_{z \in {\rm A}} |\chi^{\rm L}|^2(z)  \, ,
\end{equation}
and the bipartite entanglement entropy at late times decays as
\begin{equation}
S_{L/2}^{(2)}(t \to \infty) \simeq 2 e^{2 \overline E'' t} e^{\beta \mu} \sum_{z} \sum_{z' \in {\rm s}} |\chi^{\rm L}|^2(z \in {\rm A}) |\chi^{\rm R}|^2(z') \, .
\end{equation}
In systems of delocalized bosons, such as \eqref{dimapp}, the summations above produce a volume-independent contribution in $L \to \infty$ limit. Thus, the purification due to the dispersion in $E''(k)$ destroys the volume law scaling expected from a thermal mixture, and the late-time bipartite R\'enyi entropy does not scale with the size of the system, while approaching zero with time.\\

\underline{\textit{Phase III: $E''(k) = E''$ flat}} (non-purifying):

In this phase, all eigenmodes of the Hamiltonian survive at late times. The bipartite late-time entanglement entropy takes the form 
\begin{equation}
S_{L/2}^{(2)}(t \to \infty) \simeq 2 e^{2 E'' t} e^{\beta \mu} \sum_{E_1 E_2} \sum_{z \in {\rm A}} \sum_{z' \in {\rm s}} e^{-i (E_1'-E_2') t} \psi_{E_1}^{\rm R}(z') \psi_{E_1}^{\rm L*}(z) \psi_{E_2}^{\rm L}(z) \psi_{E_2}^{\rm R*}(z') \, .
\end{equation}
Taking into account the translational invariance of the wavefunctions, we estimate that the scaling of the entanglement entropy in this regime is at least extensive. This extensive scaling arises from the eigenmode summation, reflecting the preservation of the thermal mixture at late times. Thus, the considered model \eqref{dimapp} with thermal initial conditions exhibits a transition from purifying to a non-purifying regime when the imaginary spectrum becomes exactly flat. We confirm the scaling results above by numerically computing the entanglement entropy \eqref{sg}, \eqref{gkons} in the dimer chain model \eqref{dimapp}, sketching the initial setup, the resulting phase diagram at late times, and entanglement scaling in both phases in \cref{appfig2}.

\section{Non-Hermitian systems with Fock initial conditions}
\label{appe}

Keldysh field theory is traditionally applied to systems with Gaussian initial conditions, such as thermal density matrices and squeezed states. It is nevertheless possible to use the Keldysh path integral to describe the time evolution of systems prepared in non-Gaussian initial conditions, such as on-site Fock states. To achieve this, the contribution from the initial conditions must be exponentiated using a derivative trick \cite{Chakraborty_2019}. In particular, for $n_1,\ldots,n_L$ bosons ($\sum_x n_x = N$) initialized on sites $1,\ldots,L$, the initial density matrix contributes the following term to the partition function \eqref{parf}:
\begin{equation}
\begin{gathered}
\braket{\{\phi_{1}(1), \ldots, \phi_1(L)\}|n_1, \ldots, n_L}
\braket{n_1, \ldots, n_L|\{\phi_{2T}(1), \ldots, \phi_{2T}(L)\}}  \\
= \frac{(\overline\phi_{1}(1)\phi_{2T}(1))^{n_1}}{n_1!}\cdots
\frac{(\overline\phi_{1}(L)\phi_{2T}(L))^{n_L}}{n_L!}
= \left. \partial_{\{u_x\}} e^{\sum_x u_x \overline\phi_{1}(x)\phi_{2T}(x)} \right|_{\{u_x\} = 0}\, ,
\end{gathered}
\end{equation}
where
\begin{equation}
\partial_{\{u_x\}} = \prod_{x=1}^L \frac{1}{n_x!}\frac{\partial^{n_x}}{\partial u_x^{n_x}}\, .
\end{equation}
This construction allows one to use expressions obtained for Gaussian systems, such as~\eqref{chiRRfin}, provided that the derivatives with respect to the variables $\{u_x\}$ are evaluated afterward. For systems with Fock initial conditions (see~\eqref{gm1} , \eqref{spmtherm}) we identify
\[
\hat \Delta(x,y) = u_x \delta_{xy}\, ,
\]
which yields the trace of the density matrix as a function of time:
\begin{equation}
\operatorname{Tr}\hat{\rho}(t)
= \left.\partial_{\{u_x\}}\det\nolimits^{-1}\!\left( \hat I - \hat{\cal {G}}_{\rm s + \overline s} \hat{\Delta} \right)\right|_{\{u_x\}=0} \, .
\label{rhodif}
\end{equation}

To evaluate the derivatives, it is convenient to use MacMahon's master theorem, which expresses the inverse determinant of a matrix $\hat I - \hat M \hat \Delta$ as a sum over matrix permanents:
\begin{equation}
\det\nolimits^{-1}\!\left(\hat I - \hat M \hat{\Delta} \right)
= \sum_{n_1,\ldots,n_L} {\rm perm}\, \hat M[\{n_x\}]\prod_x \frac{u_x^{n_x}}{n_x!}\, .
\label{macmahon}
\end{equation}
Here, $\hat M[\{n_x\}]$ is an $N\times N$ matrix in which each row and column labeled by $x$ is repeated $n_x$ times. For example, $(\hat M[\{L,0,\ldots,0\}])_{ij} = M_{11}$, while $\hat M[\{1,1,\ldots,1\}] = \hat M$. The matrix permanent entering~\eqref{macmahon} is analogous to the determinant, except that all permutations enter with a positive sign:
\begin{equation}
{\rm perm}\, \hat M
= \sum_{\sigma \in {\cal S}_N} \prod_{i=1}^N M_{i\sigma(i)}\, ,
\end{equation}
to be compared with
\begin{equation}
{\rm det}\, \hat M
= \sum_{\sigma \in {\cal S}_N} (-1)^{{\rm sgn}\,\sigma} \prod_{i=1}^N M_{i\sigma(i)}\, ,
\end{equation}
where ${\cal S}_N$ is the set of all permutations of integers from 1 to $N$.

By inverting the Keldysh action as demonstrated in~\cref{appd}, a construction analogous to~\eqref{rhodif} can be obtained for the Wigner characteristic function:
\begin{equation}
\chi^{\rm W}(\{\xi_x\}, t)
= \left.\partial_{\{u_x\}}
\det\nolimits^{-1}\!\left( \hat I - \hat{\cal G}_{\rm s + \overline s} \hat{\Delta} \right)
\exp\!\left(
- \frac{1}{2} \sum_{xy} \xi_x^*
\left[
\hat I +2 \hat G^{>} \hat \Delta
(\hat I-\hat{\cal G}_{\rm s + \overline s} \hat \Delta)^{-1}
\hat G^{>\dagger}
\right]_{xy}
\xi_y
\right)\right|_{\{u_x\}=0} .
\label{chidif}
\end{equation}

The expressions for the trace of the density matrix~\eqref{rhodif} and the Wigner characteristic function~\eqref{chidif} constitute the necessary ingredients for determining the second R\'enyi entanglement entropy~\eqref{Sfin} in any non-Hermitian system with Fock initial conditions. Below, we consider two illustrative examples: first, $N$ bosons initialized on a single site $x_0$, and second, $N$ bosons distributed across $N$ distinct sites with one boson per site.

\subsection{$N$ bosons on a single site $x_0$}

In this case, we retain only a single source, such that
\[
\Delta(x,y) = u_{x_0} \delta_{x x_0} \delta_{y x_0}\, , \qquad \partial_{\{u_x\}} = (N!)^{-1}\, \partial_{u_{x_0}}^{N}\, .
\]
From Eqs.~\eqref{rhodif}, we obtain
\begin{equation}
\operatorname{Tr}\hat{\rho}(t)
= \left[{\cal G}_{\rm s + \overline s}(x_0,x_0)\right]^N\, .
\end{equation}
We now compute the Wigner characteristic function~\eqref{chidif}. The prefactor reads
\begin{equation}
\begin{aligned}
\det\nolimits^{-1}\!\left( \hat I - \hat{\cal G}_{\rm s + \overline s} \hat{\Delta} \right)
&= \exp\!\left[ - \operatorname{Tr} \ln \!\left( \hat I - \hat{\cal G}_{\rm s + \overline s}\hat{\Delta} \right) \right] \\
&= \exp\!\left[
\operatorname{Tr}\!\left( \hat{\cal G}_{\rm s + \overline s} \hat{\Delta} \right)
+ \frac{1}{2} \operatorname{Tr}\!\left( \hat{\cal G}_{\rm s + \overline s} \hat{\Delta}
\hat{\cal G}_{\rm s + \overline s} \hat{\Delta} \right)
+ \ldots
\right] \\
&= \exp\!\left[
\ln\!\left( \left[ 1 - u_{x_0} {\cal G}_{\rm s + \overline s}(x_0, x_0) \right]^{-1} \right)
\right]
= \frac{1}{1 - u_{x_0} {\cal G}_{\rm s + \overline s}(x_0,x_0)} \, .
\end{aligned}
\end{equation}
Analogously simplifying the expression in the exponent of~\eqref{chidif}, we find
\begin{equation}
\chi^{\rm W}(\{\xi_x\}, t)
= \left.
\partial_{\{u_x\}}
\frac{1}{1 - u_{x_0} {\cal G}_{\rm s + \overline s}(x_0,x_0)}
\exp\!\left(
- \frac{1}{2} \sum_{xy} \xi_x^*
\left[
\delta_{xy}
+ 2\hat{G}^{>}(x,x_0)
\frac{u_{x_0}}{1 - u_{x_0} {\cal G}_{\rm s + \overline s}(x_0,x_0)}
\hat G^{>\dagger}(x_0,y)
\right]
\xi_y
\right)
\right|_{\{u_x\}=0} \, .
\end{equation}

The derivatives with respect to $u_{x_0}$ generate the $N$th Laguerre polynomial,
\begin{equation}
\chi^{\rm W}(\{\xi_x\}, t)
= \left[{\cal G}_{\rm s + \overline s}(x_0,x_0)\right]^N
\, L_N \!\left(\sum_{xy} \xi_x^* v_x v_y^* \xi_y \right) e^{-\frac{1}{2}\sum_x |\xi_x|^2}\, ,
\qquad
v_x = \frac{G^>(x,x_0)}{\sqrt{{\cal G}_{\rm s + \overline s}(x_0,x_0)}}\, .
\end{equation}

The R\'enyi-$2$ entropy~\eqref{Sfin} then takes the form
\begin{equation}
e^{-S_{\mathrm{s}}^{(2)}}
= \int \left[\prod_y \frac{d \xi_y^*\, d \xi_y}{\pi}\right]
e^{-\sum_x |\xi_x|^2}
L_N^2\!\left(\sum_{xy} \xi_x^* \xi_y v_x v_y^* \right) .
\end{equation}
The integral over $\{\xi_x\}$ can be evaluated using
\begin{equation}
L_N(x) = \sum_{k=0}^N \binom{N}{k}\frac{(-x)^k}{k!}\, ,
\end{equation}
together with
\begin{equation}
\int \prod_x \left[\frac{d \xi_x^*\, d \xi_x}{\pi}\right]
\left(\sum_{xy} \xi_x^* \xi_y v_x v_y^* \right)^M e^{-\frac{1}{2}\sum_x |\xi_x|^2}
= M!\, \zeta^M\, ,
\qquad
\zeta = \sum_{x \in {\rm s}} |v_x|^2\, .
\end{equation}
This yields
\begin{equation}
e^{-S_{\mathrm{s}}^{(2)}}
= \sum_{k=0}^N \sum_{m=0}^N
\binom{N}{k}\binom{N}{m}
(-1)^{k+m}\frac{(k+m)!}{k!\, m!}\,
\zeta^{k+m}
= \sum_{k=0}^N \sum_{m=0}^N
\binom{N}{k}\binom{N}{m}\binom{k+m}{k}
(-\zeta)^{k+m} .
\label{SNon1}
\end{equation}
This expression is the entanglement entropy of a Bose-Einstein condensate (BEC): since all bosons are initially placed on a single site, and the non-Hermitian, non-interacting dynamics is non-scrambling, all bosons remain in the same state. As we show below, the expression \eqref{SNon1} saturates at the asymptotic value $S^{(2)} \sim \log N$ for large $N$. A similar expression for von Neumann entropy of a BEC was found in \cite{yang}.

\subsection{1 boson per each of $N$ sites}

Suppose all $N$ sites in region A initially contain one boson. Thus,
\[
\Delta(x,y)=u_x\delta_{xy}, \qquad 
\partial_{\{u_x\}}=\prod_{x \in {\rm A}} \partial_{u_x}.
\]
According to Eq.~\eqref{macmahon}, the trace of the density matrix \eqref{rhodif}
reduces to the permanent of the Green's function,
\begin{equation}
\operatorname{Tr}\hat\rho(t)=\operatorname{perm}\,[\hat{\cal G}_{\rm s + \overline s}(x,y)]_{x,y \in {\rm A}} \, .
\end{equation}

We bypass taking the derivatives with respect to $\{u_x\}$ in the expression
for the WCF \eqref{chidif}, plugging it into the expression for the R\'enyi
entropy \eqref{chiRRfin} and evaluating the Gaussian integrals over
$\{\xi_x\}$ first, and applying Sylvester's theorem
\begin{equation}
\det(\hat I-\hat A\hat B)=\det(\hat I-\hat B\hat A)\, ,
\end{equation}
which yields
\begin{equation}
\begin{aligned}
e^{-S_{\rm s}^{(2)}}=
[\operatorname{Tr}\hat\rho(t)]^{-2}
\partial_{\{u_x, \widetilde u_x\}}
&\det\nolimits^{-1}\!\left(\hat I-\hat{\cal G}_{\rm s + \overline s}\hat\Delta\right)
\det\nolimits^{-1}\!\left(\hat I-\hat{\cal G}_{\rm s + \overline s}\hat{\widetilde\Delta}\right) \\
&\left.
\times
\det\nolimits^{-1}\!\Bigl(
\hat I+
\hat {\cal G}_{\rm s}
\bigl[
\hat\Delta(\hat I-\hat{\cal G}_{\rm s + \overline s}\hat\Delta)^{-1}
+\hat{\widetilde\Delta}(\hat I-\hat{\cal G}_{\rm s + \overline s} \hat{\widetilde\Delta})^{-1}
\bigr]
\Bigr)
\right|_{\{u_x\}=\{\widetilde u_x\}=0} \, ,
\end{aligned}
\label{detdetdeeet}
\end{equation}
where $\widetilde\Delta(x,y)=\widetilde u_x\delta_{xy}$, and we introduced the restricted overlap matrix
\begin{equation}
\hat {\cal G}_{\rm s}(x,y) = \sum_{z \in \rm s} G^{>\dagger}(x,z) G^{>}(z,y)\, .
\end{equation}
The three determinants in \eqref{detdetdeeet} can be combined into the determinant of a $2N \times 2N$ matrix, as follows:
\begin{equation}
e^{-S_{\rm s}^{(2)}}=
[\operatorname{Tr}\hat\rho(t)]^{-2}
\partial_{\{u_x\}}\partial_{\{\widetilde u_x\}} \left.
\times
\det\nolimits^{-1}\!
\begin{pmatrix}
\hat I+ \hat {\cal G}_{\overline{\rm s}} \hat \Delta & \hat {\cal G}_{{\rm s}} \hat{\widetilde \Delta}  \\
\hat {\cal G}_{{\rm s}} \hat \Delta  & \hat I + \hat {\cal G}_{\overline{\rm s}} \hat{\widetilde \Delta}
\end{pmatrix}
\right|_{\{u_x\}=\{\widetilde u_x\}=0} \, ,
\label{master}
\end{equation}
The result of differentiation is determined using MacMahon's theorem \eqref{macmahon}, which yields the final expression for the R\'enyi-2 entropy of $N$ bosons uniformly distributed over $N$ sites:
\begin{equation}
e^{-S_{\rm s}^{(2)}}=
\frac{
\operatorname{perm}
\begin{bmatrix}
{\cal G}_{\rm s}(x,y)
&
{\cal G}_{\overline{\rm s}}(x,y)
\\
{\cal G}_{\overline{\rm s}}(x,y)
&
{\cal G}_{\rm s}(x,y)
\end{bmatrix}_{x,y \in {\rm A}}
}{
\left(\operatorname{perm} \left[\hat{\cal G}_{\rm s + \overline s} (x,y)\right]_{x,y \in {\rm A}}\right)^2
} .
\label{SNN}
\end{equation}
This expression can be reproduced from operator algebra by computing the R\'enyi entropy from 
\begin{equation}
e^{- S_{\rm s}^{(2)}}  = \frac{{\rm Tr} \left[ \, \left( \bra{\psi(t)} \otimes \bra{\psi(t)} \right) {\rm Swap}_{\overline{\rm s}} \left( \ket{\psi(t)} \otimes \ket{\psi(t)} \right) \right]}{\braket{\psi(t) | \psi(t)}^2}\, ,
\label{SNQM}
\end{equation}
where the swap operator acts only on a complement $\overline{\rm s}$, interchanging the bosonic states in the first and second replicas, and the state $\ket{\psi(t)}$ results from the time evolution with a non-unitary Hamiltonian $\hat H$
\begin{equation}
\ket {\psi(t)} = e^{- i \hat H t}  \prod_{x \in {\rm A}} \hat b^\dagger_x  \ket{0} = \prod_{x \in {\rm A}} \sum_y G^>(y,x) \hat b^\dagger_y \ket{0} \, .
\label{psitiQM}
\end{equation}
The permanents in \eqref{SNN} then follow from \eqref{psitiQM} and \eqref{SNQM} as a result of the indentity
\begin{equation}
\braket{0| \prod_{\tilde x \in {\rm A}} \sum_{\tilde y \in {\rm s}} \hat b_{\tilde y}  G^{>\dagger}(\tilde x, \tilde y) | \prod_{x \in {\rm A}} \sum_{y \in {\rm s}} G^>(x,y) \hat b^\dagger_y |0} = {\rm perm} \left[\hat {\cal G}_{\rm s}(x,y)\right]_{x,y \in {\rm A}} \, .
\end{equation}

We note that the expression \eqref{SNN} is a non-Hermitian generalization of the known result~\cite{kaga_2023}, reducing to the latter in the Hermitian limit, where $\hat {\cal G}_{\rm s+ \overline s}=\hat I$. Since the non-unitary evolution neither creates nor destroys particles, the instantaneous eigenstate remains pure. Indeed, for ${\rm s}=L$,
\begin{equation}
e^{-S_{L}^{(2)}}=
\frac{
\operatorname{perm}
\begin{bmatrix}
\hat{\cal G}_{\rm s + \overline s}(x,y) & 0 \\
0 & \hat{\cal G}_{\rm s + \overline s}(x,y)
\end{bmatrix}_{x,y \in {\rm A}}
}{
\left(\operatorname{perm} \left[\hat{\cal G}_{\rm s + \overline s}(x,y) \right]_{x,y \in {\rm A}}\right)^2
}
=1 .
\end{equation}

The permanent expression \eqref{SNN} has a complicated structure, involving
many independently oscillating in time contributions. Since the computational complexity of permanents scales exponentially with the number of particles, systems with $N\gtrsim20$ are already numerically challenging. However, if the Liouvillian spectrum $E''(k)$ is dispersive and features a well-defined maximum, only a single mode $\chi(x)$ in the Green’s function summations \eqref{Gg}, \eqref{Gt} survives at $t\to\infty$, yielding
\begin{equation}
e^{-S_{\rm s}^{(2)}}=
\frac{
\operatorname{perm}
\begin{bmatrix}
c_{\rm s}^{\rm R}\chi^{\rm L}(x)\chi^{\rm L*}(y)
&
(c^{\rm R}-c_{\rm s}^{\rm R})\chi^{\rm L}(x)\chi^{\rm L*}(y)
\\
(c^{\rm R}-c_{\rm s}^{\rm R})\chi^{\rm L}(x)\chi^{\rm L*}(y)
&
c_{\rm s}^{\rm R}\chi^{\rm L}(x)\chi^{\rm L*}(y)
\end{bmatrix}_{x,y \in {\rm A}}
}{
\left(
\operatorname{perm}
\left[
c^{\rm R}\chi^{\rm L}(x)\chi^{\rm L*}(y)
\right]_{x,y \in {\rm A}}
\right)^2
} .
\label{Ssing}
\end{equation}

Here
\[
c^{\rm R}=\sum_x |\chi^{\rm R}(x)|^2,
\qquad
c_{\rm s}^{\rm R}=\sum_{x\in{\rm s}} |\chi^{\rm R}(x)|^2 .
\]

The denominator simplifies to
\begin{equation}
\operatorname{perm}
\bigl[
c^{\rm R}\chi^{\rm L}(x)\chi^{\rm L*}(y)
\bigr]
=(c^{\rm R})^N N! \prod_{x \in {\rm A}} |\chi^{\rm L}(x)|^2 .
\end{equation}

The numerator can be treated analogously, which gives
\begin{equation}
e^{-S_{\rm s}^{(2)}}=
\sum_{m=0}^N
\binom{N}{m}^2
\left(\frac{c_{\rm s}^{\rm R}}{c^{\rm R}}\right)^{2m}
\left(1-\frac{c_{\rm s}^{\rm R}}{c^{\rm R}}\right)^{2N-2m}
=
\sum_{k,m=0}^N
\binom{N}{k}\binom{N}{m}\binom{k+m}{k}
\left(-\frac{c_{\rm s}^{\rm R}}{c^{\rm R}}\right)^{k+m} \, .
\label{sbinomial}
\end{equation}

This expression represents the entanglement entropy of a Bose–Einstein condensate occupying the slowest decaying mode $\chi$ and is analogous to Eq.~\eqref{SNon1}. Thus, when only a single mode survives in the system at $t \to \infty$, the entanglement entropy scales logarithmically with the volume. This can be explicitly verified by letting $c_{\rm s}^{\rm R}/c^{\rm R}  = 1/2$ (the dark mode occupies exactly 1/2 of the system), which gives in $N \to \infty$ limit
\begin{equation}
S_{L/2}^{(2)}= - \log \left[
\sum_{m=0}^N
\binom{N}{m}^2
(1/2)^{2N} \right] = - \log \left[ \binom{2N}{N} (1/2)^{2N} \right] \simeq - \log \left[ (\pi N)^{-1/2} \right] = \frac{1}{2} \log (\pi N) \, .
\end{equation}

Lastly, we note that while we consider the entanglement entropy of the system initialized in the singly occupied Fock states here, MacMahon's master theorem \eqref{macmahon} can be used to evaluate \eqref{master} in terms of permanents with any set of auxiliary derivatives, including those corresponding to degenerate initial bosonic occupations.

\subsection{Mixture of on-site Fock and thermal initial conditions}
\label{appf}

The formalism introduced above also allows the determine the entanglement entropy in systems initialized in mixtures of thermal and Fock states, as we show below. Consider the density matrix
\begin{equation}
\hat \rho_{\rm mix} = p \, \hat {\overline\rho}_{\rm th} + (1-p) \, \hat \rho_{N} \, ,
\label{rhomix}
\end{equation}
where $\hat {\overline\rho}_{\rm th} = \hat {\widetilde\rho}_{\rm th}/{\widetilde \rho}$, ${\widetilde\rho} = {\rm Tr}\, \hat {\widetilde\rho}_{\rm th}$ , $\hat {\widetilde\rho}_{\rm th}$ is defined in \eqref{trhth}, $\hat \rho_{N}$ corresponds to 1 boson on each site of a subsystem A considered in the previous section, and $p$ is a free parameter varying from 0 to 1. We further define
\begin{equation}
\Delta(x,y) = \delta_{x,y} u_x \, , \qquad \Delta_{\rm th}(x,y) = \delta_{x,y} e^{- \beta_x (\varepsilon_x - \mu)} \, , \qquad \partial_{\{u_x\}} = \prod_{x \in {\rm A}} \partial_{u_x} \, .
\end{equation}
We now focus on obtaining the R\'enyi entanglement entropy in a general non-Hermitian system \eqref{Sfin} initialized in $\hat \rho_{\rm mix}$ \eqref{rhomix} at time $t$. The normalization denominator is given by
\begin{equation}
{\rm Tr} \, \hat \rho_{\rm mix} = (p/{\widetilde\rho}) \, {\rm det}^{-1} ( \hat I - \hat{\cal G}_{\rm s+ \overline s}\hat \Delta_{\rm th} ) + (1-p) \, {\rm perm}\, \hat{\cal G}_{\rm s+ \overline s} \, .
\end{equation}
The Wigner characteristic function \eqref{wcfdef} is linear with respect to the initial condition, and separates into
\begin{equation}
\chi^{\rm W}_{\rm mix}(t, \{\xi_x\}) =  \chi^{\rm W}_{\rm th}(t, \{\xi_x\}) + \chi^{\rm W}_{N}(t, \{\xi_x\}) \, .
\end{equation}
The numerator term in \eqref{Sfin} can be expanded as 
\begin{equation}
\int \left \{ \prod_x \frac{d\xi_x^* d\xi_x}{\pi} \right \} \chi^{\rm W}_{\rm mix} \chi^{\rm W}_{\rm mix} = \int \left \{ \prod_x \frac{d\xi_x^* d\xi_x}{\pi} \right \} \chi^{\rm W}_{N} \chi^{\rm W}_{N} + \int \left \{ \prod_x \frac{d\xi_x^* d\xi_x}{\pi} \right \} \chi^{\rm W}_{\rm th} \chi^{\rm W}_{\rm th} +2 \int \left \{ \prod_x \frac{d\xi_x^* d\xi_x}{\pi} \right \} \chi^{\rm W}_{\rm th} \chi^{\rm W}_{N} \, ,
\end{equation}
where
\begin{equation}
\begin{aligned}
\int \left \{ \prod_x \frac{d\xi_x^* d\xi_x}{\pi} \right \} \chi^{\rm W}_{N} \chi^{\rm W}_{N} & =(1-p)^2 \operatorname{perm}
\begin{bmatrix}
{\cal G}_{\rm s}(x,y)
&
{\cal G}_{\overline{\rm s}}(x,y)
\\
{\cal G}_{\overline{\rm s}}(x,y)
&
{\cal G}_{\rm s}(x,y)
\end{bmatrix}_{x,y \in {\rm A}} \, , \\
\int \left \{ \prod_x \frac{d\xi_x^* d\xi_x}{\pi} \right \} \chi^{\rm W}_{\rm th} \chi^{\rm W}_{\rm th} & = (p/{\widetilde\rho})^2 {\rm det}^{-2} ( \hat I - \hat {\cal G}_{\rm s + \overline s}\hat \Delta_{\rm th}) \, {\rm det}^{-1} \left[ \hat I + 2 \hat {\cal G}_{\rm s} \hat \Delta_{\rm th} (\hat I-\hat {\cal G}_{\rm s + \overline s} \hat \Delta_{\rm th})^{-1} \right] \, , \\
\int \left\{ \prod_x \frac{d\xi_x^* d\xi_x}{\pi} \right\} \chi^{\rm W}_{\rm th} \chi^{\rm W}_{N}
& = (p/\widetilde{\rho}) (1-p) \, {\rm det}^{-1}
\left[ \hat I - \hat {\cal G}_{\overline{\rm s}} \hat \Delta_{\rm th} \right] \operatorname{perm}
\left[ (\hat I + \hat g_{\rm th})^{-1}
[\hat {\cal G}_{\overline{\rm s}}
+ \hat g_{\rm th} \hat{\cal G}_{\rm s + \overline s} ]\right]_{{\rm A}} \, , 
\end{aligned}
\end{equation}
and
\begin{equation}
\hat g_{\rm th} = \hat{\cal G}_{\rm s} \hat \Delta_{\rm th} (\hat I-\hat {\cal G}_{\rm s + \overline s} \hat \Delta_{\rm th})^{-1}  \, .
\end{equation}

\end{document}